\documentclass[fleqn,usenatbib]{mnras}

\usepackage{newtxtext,newtxmath}
\usepackage{anyfontsize}

\usepackage[T1]{fontenc}

\DeclareRobustCommand{\VAN}[3]{#2}
\let\VANthebibliography\thebibliography
\def\thebibliography{\DeclareRobustCommand{\VAN}[3]{##3}\VANthebibliography}

\usepackage{graphicx}	
\usepackage{amsmath}	

\title[Co-evolution of Black Holes and $L^{\star}$ Galaxies]{Co-evolution of Supermassive Black Holes and their Host $L^{\star}$ galaxies: implications for Milky Way and M31}
\author[Grimozzi et al.]{Salvador E. Grimozzi$^{1,2}$, Mar\'{\i}a Emilia De Rossi$^{2,3}$, Andreea S. Font$^{4}$\thanks{A.S.Font@ljmu.ac.uk}\\
$^{1}$Universidad de Buenos Aires, Facultad de Ciencias Exactas y Naturales, Departamento de Física. Buenos Aires, Argentina\\
$^{2}$CONICET-Universidad de Buenos Aires, Instituto de Astronomía y Física del Espacio (IAFE), Buenos Aires, Argentina\\
$^{3}$Universidad de Buenos Aires, Facultad de Ciencias Exactas y Naturales y Ciclo Básico Común, Buenos Aires, Argentina\\
$^{4}$Astrophysics Research Institute, Liverpool John Moores University, 146 Brownlow Hill, Liverpool L3 5RF, UK\\}

\date{Accepted XXX. Received YYY; in original form ZZZ}

\pubyear{2025}

\begin{document}
\label{firstpage}
\pagerange{\pageref{firstpage}--\pageref{lastpage}}
\maketitle

\begin{abstract}
We investigate the origin of the scatter in supermassive black hole (BH) masses among $L^{\star}$ galaxies, using the \texttt{ARTEMIS} and \texttt{EAGLE} simulations. We classify galaxies by their central BH-to-stellar mass ratios and track systems with the lowest and highest ratios (LBH and HBH, respectively). Although these populations have similar properties at $z\approx 2$, they diverge significantly towards lower redshifts. LBH galaxies maintain higher gas fractions and sustain star formation, while HBHs form stars earlier, experience faster BH growth, more efficient feedback, and subsequent quenching. The simulations broadly match the observed scatter in the BH masses and galaxy morphologies in the $L^{\star}$ regime, linking it to differences in merger histories. Galaxies with more active merger histories host more massive BHs at present and are predominantly elliptical, whereas those with quieter histories host lower-mass BHs and tend to be disc-like. Mergers induce BH growth through enhanced gas accretion and direct BH--BH coalescences. However, these channels operate differently: in HBH galaxies, BHs grow primarily ($\approx 90$\%) through gas accretion, whereas in LBHs they grow  through both gas accretion and BH--BH mergers, in comparable fractions. These results suggest that the contrasting BH masses in the Milky Way and M31 may arise from differences in their merger histories. We also investigate the role of feedback from supernovae (SN) versus active galactic nuclei (AGN) on these scales. AGN feedback appears to be the primary mechanism responsible for quenching MW-mass systems, although SN feedback remains the dominant regulator of their overall evolution.
\end{abstract}

\begin{keywords}
methods: numerical -- galaxies: evolution -- galaxies: formation -- galaxies: haloes -- (galaxies) quasars: supermassive black holes
\end{keywords}



\section{Introduction}
\label{sec:intro}

Supermassive black holes (SMBHs; BHs, hereafter) are ubiquitous in the Universe. They inhabit central regions of galaxies that span many orders of magnitude in mass, from the brightest and most massive galaxies in clusters to dwarf galaxies \citep{Kormendy1995,Genzel2010,Kormendy2013,Taylor2025}. The properties of BHs and their host galaxies are closely connected, as evidenced by a variety of scaling relationships, such as those between BH masses ($M_{\mathrm{BH}}$) and bulge masses ($M_{\mathrm{sph}}$), total stellar masses ($M_{\star}$), or stellar velocity dispersions ($\sigma$) \citep{Maggorian1998,Ferrarese2000,Haring2004,McConnell2013,Heckman2014,Reines2015}. The $M_{\rm BH}-\sigma$ relation is one of the tightest BH scaling relations (\citealt{Shankar2025}), which may be a consequence of $\sigma$ being correlated with $M_{\mathrm{sph}}$, or being sensitive to the depth of the potential well in which BHs form \citep{Ferrarese2000}. These observed trends suggest a strong degree of co-evolution between BHs and their host galaxies. 

The scatter in these scaling relationships may also contain information about this co-evolution. For example, in the $L^{\star}$ regime, the Milky Way (MW) has one of the lowest-mass central BHs, Sgr A*, of $\approx 4.3\times10^{6}~\mathrm{M_{\odot}}$ (\citealt{Gravity2023}), while M31 has a central BH of $\approx 1.4\times10^{8}~\mathrm{M_{\odot}}$ \citep{Bender2005}. This represents nearly two orders of magnitude difference in BH masses for comparable host stellar masses. However, M31 has an inner spheroid that is more massive than the MW bulge \citep{Reitzel1998}, and shows evidence of a massive accretion event at more recent times \citep{Ibata2001}. Therefore, central BHs may retain information about the assembly histories of galaxies, despite the vast differences in spatial scales associated with them.

The connection between the growth of central BH and galaxy properties can be  established via one (or both) of two primary channels: BH growth via smooth gas accretion or induced by galaxy mergers. Although the former channel appears to have been dominant in the early Universe, the latter becomes increasingly important at lower redshifts ($z$), particularly for massive systems (\citealt{Dubois2014,Bravo2025}). Mergers can drive gas accretion by inducing gravitational torques within the interstellar medium (ISM).  Tidal forces effectively dissipate the angular momentum of the gas, driving it into nuclear regions where it serves as fuel for the central BH (\citealt{Jogee2006,Hopkins2010}). In addition, mergers lead to the coalescence of central BHs. They also drive morphological changes by modifying the angular momentum of baryons, which helps to explain the observed correlations between galaxy morphologies and BH masses \citep{Greene2020,GrahamSahu2023a}.
Although it is generally agreed that mergers are important for BH growth, many questions remain about how this mechanism effectively operates, e.g., whether they trigger AGN feedback and what their relevance is across many scales in galaxy mass (\citealt{Ellison2019,Villforth2023,Alexander2025}).

The MW stellar mass is around the characteristic mass $M_{\star ,{\rm c}} \approx 5 \times10^{10}~\mathrm{M_{\odot}}$ where significant variations in the global properties of local galaxies are observed. For example, some studies suggest that the $M_{\mathrm{BH}}-M_{\mathrm{sph}}$ relation could be described by a double power law, with a change in slope at $\simeq M_{\mathrm{sph}}\sim 5 \times 10^{10}~\mathrm{M_{\odot}}$: 
a remarkably steeper slope is obtained for galaxies below this mass than for galaxies above it \citep[e.g.,][]{Graham2015,DavisB2019}. Below this characteristic mass, galaxies are often spirals and actively forming stars, while above they tend to be ellipticals with little or no star formation \citep{Kauffmann2003,Baldry2006,Savorgnan2016,Comparat2022,GrahamSahu2023a}. 

The strong negative AGN feedback in massive galaxies is generally considered the primary mechanism driving these changes in galaxy properties.  AGN feedback regulates the gas supply, effectively shutting down star formation and transitioning galaxies from the blue cloud to the red sequence.
Either the heating of gas by energy released from the BH or its ejection via AGN-driven outflows can suppress star formation\footnote{Under certain conditions, positive AGN feedback is also possible. For example,  star formation can be enhanced when the outflows generated by BHs compress the gas in the ISM \citep{Zinn2013,Schutte2022,Venturi2023}. Interestingly, there is evidence that both positive and negative feedback can occur simultaneously within the same galaxy (see \citealt{Cresci2015}).} in massive galaxies \citep{Croton2006,CanoDiaz2012,Bischetti2022,Chen2022,Lammers2023}.
AGN feedback affects even the chemical properties of galaxies and their circumgalactic media, phenomena that are intrinsically coupled to the evolution of gas-phase reservoirs and the broader star formation history \citep{Taylor2015,Li2024,VillarMartin2024}; in particular, AGN feedback may be responsible for the flattening of the mass--metallicity relation at $M_{\star} \gtrsim 10^{10}~{\rm M}_{\odot}$ \citep{DeRossi2017}.

MW-mass galaxies occupy a key transition regime between two major feedback regimes: star-formation-driven outflows in low-mass haloes and BH-driven quenching in massive systems.
\cite{Bower2017} showed that the baryonic flows of galaxies dramatically change when they reach the MW-mass regime, specifically that the gas buildup at the center is blocked by stellar-driven winds in any galaxy halo less massive than $\sim 10^{12}~\mathrm{M_{\odot}}$. The star formation in these galaxies is regulated by a balance between outflows triggered by supernovae (SNe) and gas infall.  In more massive galaxies, the deeper potential well allows a hot corona to form \citep{WhiteFrenk1991,Birnboim2003}. This suppresses the buoyancy of SN-driven winds, causing the outflows to stall; as gas subsequently builds up in the core, the black hole undergoes a phase of rapid, non-linear growth. By heating the halo gas, the AGN feedback halts the inflow of cold material, which quenches galaxies once they reach the mass threshold of $\sim 10^{12}~\mathrm{M_{\odot}}$. In massive galaxies, stellar and SN feedback becomes  inefficient while AGN feedback dominates. Remarkably, the bridge phase occurs around the halo mass of the MW. The interplay between SN and AGN feedback can give rise to various BH growth histories, which may explain the substantial scatter in the present-day BH properties of $L^{\star}$ galaxies \citep{Trayford2016,Terrazas2017,Terrazas2020,Habouzit2021}.  

Cosmological hydrodynamical simulations routinely incorporate AGN feedback in galaxies that match the observed properties of galaxies (e.g., \citealt{Springel2005a,Sijacki2015,Schaye2015,McCarthy2017,Nelson2018,Dave2019,Ni2022,Husko2026,Schaye2026}), with continuing progress taking place in the modelling  \citep{Alexander2012,Harrison2017,Alexander2025}. These simulations tend to cover large volumes, encompassing orders of magnitude in galaxy mass. Nevertheless, the $L^{\star}$ regime has been extensively explored, given its relevance to both SN feedback and AGN feedback. The scatter in BH masses and the correlations between BH growth and galaxy properties in this regime have begun to be explored more recently, particularly in terms of the correlations with the mass and formation times of galaxy haloes, or with the corresponding binding energy \citep{Davies2019,Davies2022,Davies2024,Roberts2026}. 

In this study, we focus on the origin of the scatter in the central black hole mass -- galaxy stellar mass relation in the $L^{\star}$ regime. For this, we use \texttt{ARTEMIS}, a suite of zoomed-in, cosmological simulations of $42$~MW-mass galaxies \citep{Font2020}, and samples of MW-mass galaxies extracted from a series of runs from the \texttt{EAGLE} project \citep{Schaye2015,Crain2015}. We therefore assemble a large sample of MW-mass galaxies with diverse merger histories that have formed under different assumptions of SN and AGN feedback.  The \texttt{ARTEMIS} simulations successfully reproduce the main observed properties of MW analogues and their satellites \citep{Font2021,Font2022, Grimozzi2024, Grimozzi2025}, while \texttt{EAGLE} simulations match a wide range of observables for galaxies on both small and large scales \citep{Schaye2015}. The physical prescriptions in these two simulations are very similar, with differences that are complementary in nature.  The \texttt{EAGLE} simulations model larger cosmological volumes and so contain many more galaxies in the MW-mass range. They also include runs with variations of SN and AGN feedback, which will also be analysed here. \texttt{ARTEMIS} has higher resolution and can better resolve smaller merger events. Therefore, a comparison between these two simulations allows us to examine in more detail the interplay of physical processes and mergers in generating the scatter in BH masses at MW-mass scale.

This paper is organized as follows. In Section~\ref{sec:simulations}, we summarize the main characteristics of the \texttt{ARTEMIS} and \texttt{EAGLE} simulations. In Section~\ref{sec:Results}, we analyze  correlations between current BH masses and key host galaxy properties (\S~\ref{sec:BH_sr}), and study the evolutionary paths of galaxy populations at both ends of the scatter in BH masses (\S~\ref{sec:Evolution_properties}) or in stellar spins, which are proxies for morphology (\S~\ref{sec:Morphology}). We also investigate whether the trends obtained from simulations can be explained by differences in the merger histories, by focusing on two galaxies that are representative of the two BH-mass populations (\S~\ref{sec:Merger_Histories}). We discuss our results in Section~\ref{sec:discussion} and present our conclusions in Section~\ref{sec:concl}.


\section{\texttt{ARTEMIS} and \texttt{EAGLE} simulations}
\label{sec:simulations}

\texttt{ARTEMIS} is a suite of zoomed-in, cosmological hydrodynamical simulations of MW-mass galaxies. The baryons are followed with a numerical resolution that is 8 times higher than in the \texttt{EAGLE} Recal ($25~\mathrm{cMpc}\,h^{-1}$) simulations\footnote{Length units prefixed with `c' (e.g., cMpc) refer to comoving distances, and those prefixed with `p' (e.g., pkpc) to proper distances.}, specifically dark matter particle masses of $1.17\times10^5$ M$_{\sun}\,h^{-1}$, and initial gas particle masses of $2.23\times10^4$ M$_{\sun}\,h^{-1}$. The force resolution (i.e., the Plummer-equivalent softening) is $125$~pc $h^{-1}$. \texttt{ARTEMIS} uses the same hydrodynamical code as in the \texttt{EAGLE} simulations, and the same subgrid parameters as in \texttt{EAGLE} Recal model, apart from an additional re-calibration of parameters related to supernova feedback. A flat $\Lambda\mathrm{CDM}$ WMAP9 cosmological model (\citealt{Hinshaw2013}) is adopted, with the following parameters: $\Omega_\textrm{m}=0.2793$, $\Omega_\textrm{b}=0.0463$, $h=0.70$, $\sigma_8=0.8211$ and $n_s=0.972$.  The sample of 42 MW-mass galaxies used here have present-day virial masses in the range of $6.7\times10^{11}< M_{200,\mathrm{crit}}/\mathrm{M_{\odot}}<3.6\times10^{12}$ \citep{Font2020}.

The \texttt{EAGLE} suite consists of several cosmological hydrodynamical simulations run with a modified version of the \texttt{GADGET-3} code, with periodic boundary conditions. The comoving side-lengths of various cubic boxes range between $12~\mathrm{cMpc}\,h^{-1}$ and $100~\mathrm{cMpc}~h^{-1}$, including different levels of resolution (see \citealt{Schaye2015}, \citealt{McAlpine2016}, for details).   
The assumed cosmological parameters are: $\Omega_{\rm{m}}=0.307$, $\Omega_{\Lambda}= 0.693$, $\Omega_{\rm{b}}=0.04825$, $h=0.6777$, $\sigma_{8}=0.8288$, $n_{\rm{s}}=0.9611$ \citep{Planck2014}.

For this work, we use mainly the runs with box side length $L = 50~{\rm cMpc}$ and an initial number of particles per species of $N = 752^3$ (i.e., the `L050N0752' model and its variants). The reason is twofold: firstly, the large cosmological volume allows us to assemble a statistically representative sample of MW-mass galaxies, and secondly, these runs include variations of AGN feedback that are relevant for our analysis; we also study a set of simulations with $L = 25~{\rm cMpc}$ and $N = 376^3$, which include variations of SN feedback (results of these models are presented in the Appendix~\ref{sec:appendix}). The $50~{\rm cMpc}$ runs adopt initial masses of $1.81\times10^{6}~\mathrm{M_{\odot}} \, h^{-1}$ for gas particles and masses of $9.70\times10^{6}~\mathrm{M_{\odot}}\, h^{-1}$ for dark matter particles.  From the reference \texttt{EAGLE} run (described below), we selected a sample of 201 MW-mass galaxies; these are central galaxies with present-day virial masses in the range of $6.7\times10^{11}<M_{200,\mathrm{crit}}/\mathrm{M_{\odot}}<3.6\times10^{12}$, in the same range as the \texttt{ARTEMIS} haloes. In the following, we outline the main subgrid prescriptions in these two simulation suites.  

\subsection{Subgrid models}
\label{sec:subgrid_models}

Physical processes that cannot be resolved in the simulations are implemented via subgrid models. 
\texttt{ARTEMIS} applies the same subgrid prescriptions developed for the \texttt{EAGLE} project, described in more detail in \cite{Crain2015} and \cite{Schaye2015}. These include  metal-dependent radiative cooling in the presence of a photo-ionizing UV background, star formation, stellar and chemical evolution, formation of BHs, stellar feedback from SN and stellar winds, and AGN feedback.

Star formation is modeled with the prescription of \citet{Schaye2008}, where  gas with a density above $n_H = 10^{-1}\, \rm {cm}^{-3}$ (the critical density for onset of the thermo-gravitational instability in hydrogen gas) can form stars. An effective equation of state for the cold gas is also imposed in the subgrid physics. SNe inject energy into the ISM using the kinetic feedback prescription of \citet{DallaVecchia2008}. The chemical evolution is tracked independently for 11 elements (H, He, C, N, O, Ne, Mg, Si, Fe, S and Ca), whose abundances depend on the nucleosynthetic yields of AGB stars, stellar winds and core collapse SNe \citep{Wiersma2009}. 
To calculate the mass loss and the resulting chemical composition, the model combines the nucleosynthetic yields of \cite{Marigo2001} and \cite{Portinari1998} with the fraction of stellar mass in each particle reaching the end of the main sequence. The feedback and mass loss from SNIa are implemented separately, using the yields of \cite{Thielemann2003}. 

The black hole growth model is based on the \citet{Springel2005b} formalism, with subsequent refinements from \cite{Booth2009} and \cite{RosasGuevara2015}. Black holes are seeded with a mass of $10^{5}~\mathrm{M_{\odot}}\,h^{-1}$ in haloes with a total mass above $10^{10}~\mathrm{M_{\odot}}\,h^{-1}$, by replacing the densest gas particle with a collisionless black hole particle. Black holes can then grow by accretion of gas or by merging with other black holes. The gas accretion rate determines the rate at which energy is thermally injected into the ISM. A fraction of the rest-mass energy of the gas is released, powering the AGN feedback. Black holes accumulate energy at a rate that is proportional to their gas-mass accretion rate, thus increasing the feedback energy reservoir. Once a certain threshold is reached, the AGN feedback operates by stochastically heating the neighbouring particles. This threshold is equal to the amount of energy required to increase the temperature of a gas particle by $\Delta T_{\mathrm{AGN}}$. When galaxies merge, their central BHs also merge, provided that they meet certain distance and velocity criteria.

In the \texttt{EAGLE} reference (`Ref-')model, the subgrid parameters were calibrated to improve the match to the observed $z \approx 0$ galaxy stellar mass function, the stellar mass--size relation, and the normalization of the $M_{\rm BH}-M_\star$ relation. The \texttt{ARTEMIS} simulations use the \texttt{EAGLE} recalibrated model (`Recal-L025N0752'; $L=25~\mathrm{cMpc}~h^{-1}$, $N=752^{3}$), which improves the matching with the observations on small scales compared to \texttt{EAGLE} reference model. In \texttt{ARTEMIS}, the efficiency of SN feedback is further calibrated to better match the stellar mass -- halo mass relation around $L^{\star}$ \citep{Moster2018, Behroozi2019}. The \texttt{EAGLE} `Ref-' model uses $\Delta T_{\mathrm{AGN}}=10^{8.5}~\mathrm{K}$, while \texttt{ARTEMIS} uses $\Delta T_{\mathrm{AGN}}=10^{9}~\mathrm{K}$. For more details on the subgrid models, the reader is referred to \citet{Schaye2015}.

\begin{figure}
\includegraphics[width=\columnwidth]{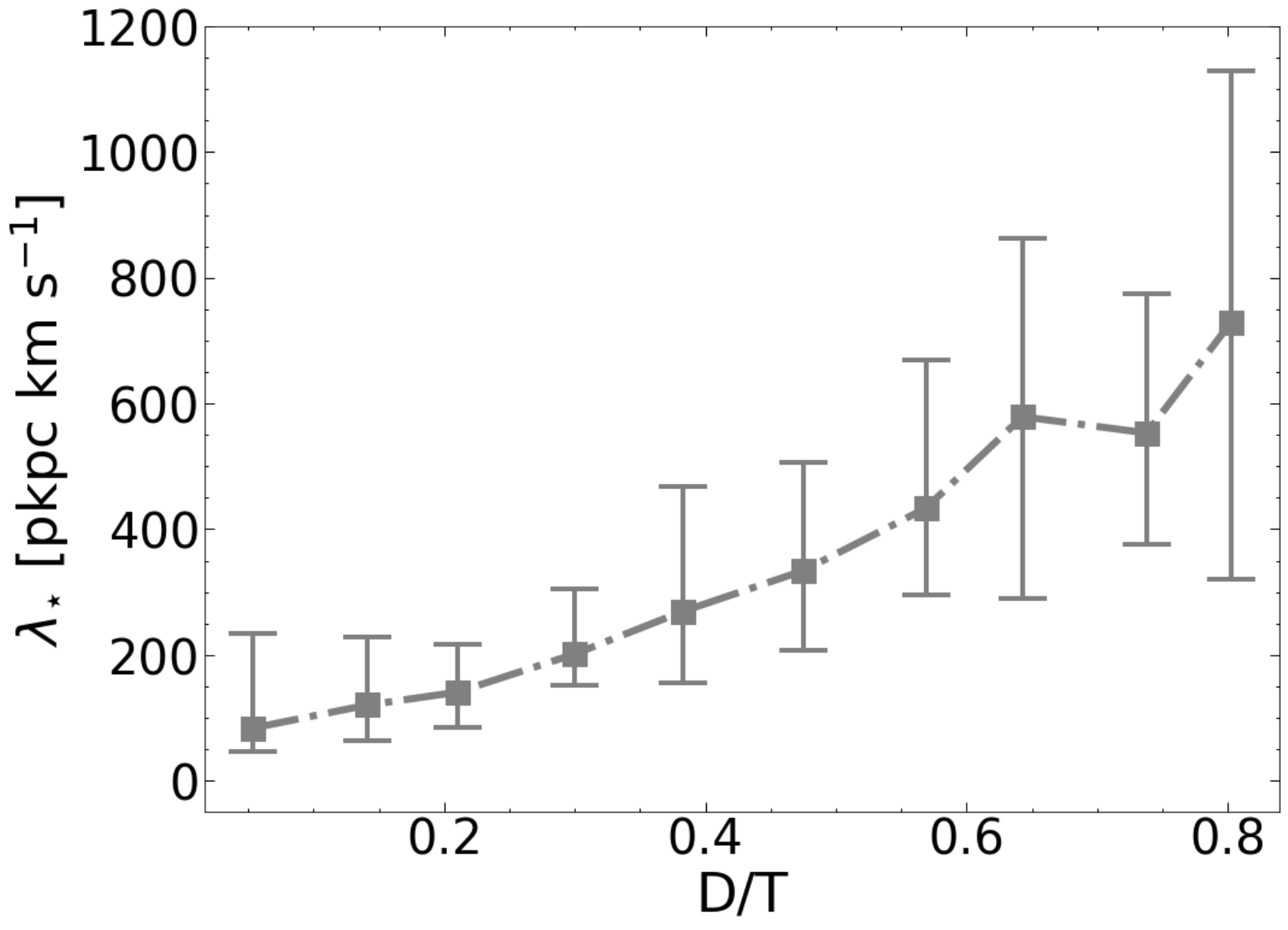}
\caption{Stellar spin ($\lambda_{\star}$) versus disc-to-total mass fraction, $D/T$, computed by \citealt{Thob2019}), for galaxies with $M_{\star}>10^{9.5}~\mathrm{M_{\odot}}$ in the \texttt{EAGLE} Recal-L025N0752 run.  The dashed line represents the median relation and the error bars the respective 25$^{\rm th}$ and 75$^{\rm th}$ percentiles.}
\label{fig:Stars_spin_vs_D_T_Recal}
\end{figure}

\begin{figure*}
\includegraphics[width=2\columnwidth]{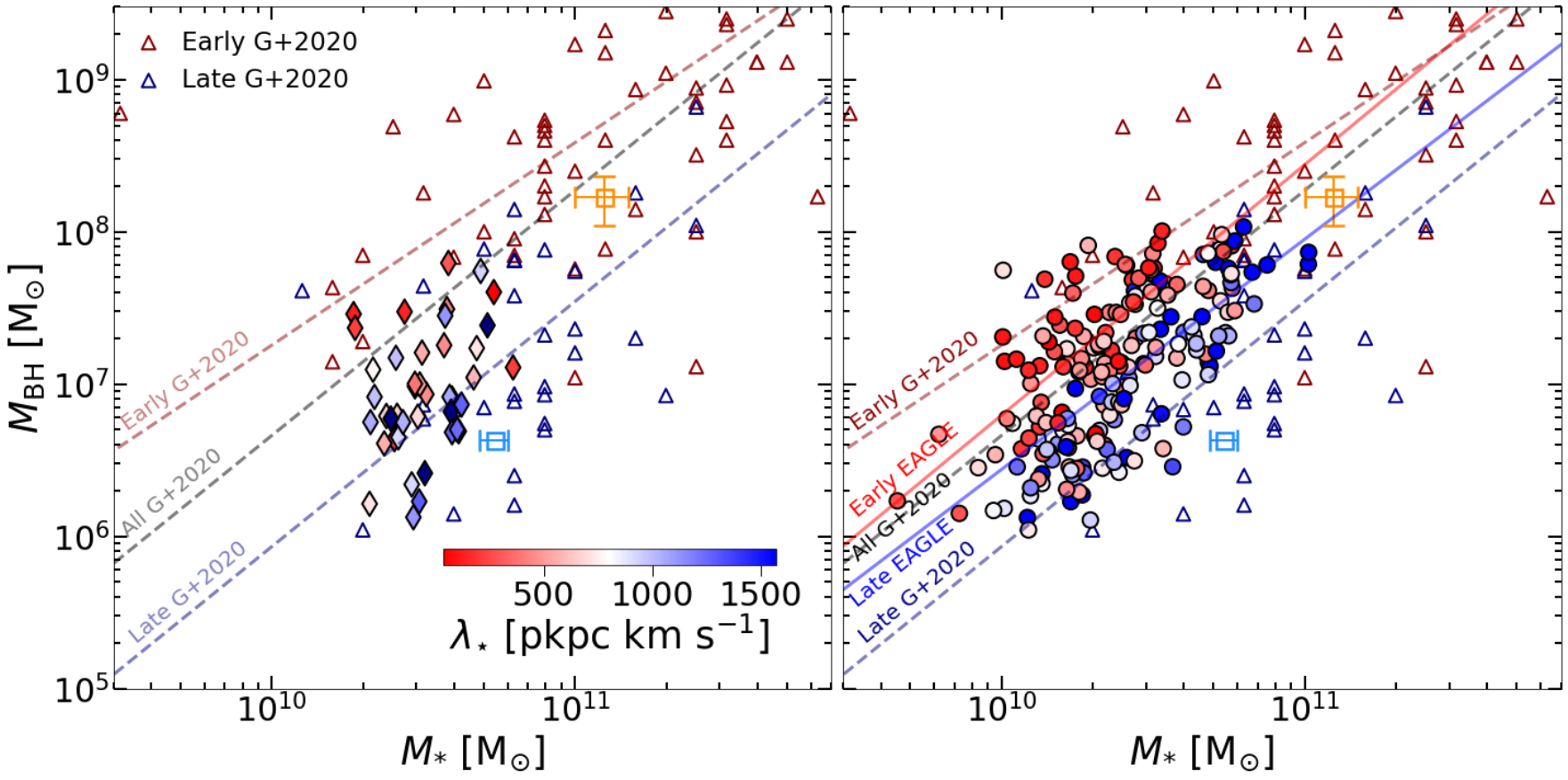}
\caption{$M_{\mathrm{BH}}-M_{\star}$ relation for simulations versus observations.  \texttt{ARTEMIS} galaxies are shown in the left panel, with  diamond symbols, and \texttt{EAGLE} galaxies on the right, with circles. Observational data from \citet{Greene2020} (and references therein) are shown with open triangles, coloured red for early-type and blue for late-type galaxies. The symbols for simulated galaxies are colour-coded by their stellar spin, $\lambda_{\star}$. The dashed lines indicate the linear fits derived from observations, for early- (red), late-type (blue) and all (gray) galaxies. Solid lines in the right panel represent the linear fits to \texttt{EAGLE} galaxies, for 
systems with low (red) and high (blue) $\lambda_{\star}$, using $350~\mathrm{pkpc~km~s^{-1}}$ as a threshold to separate morphologies. The light blue and orange squares in both panels are the measurements for MW \citep{McMillan2017,Gravity2023} and M31 \citep{Bender2005,Tamm2012}.} 
\label{fig:Mstar_MBH_relation}
\end{figure*}

\subsection{Galaxy catalogues}
\label{sec:SampAndDef}

All parameters used in this work come from the public \texttt{EAGLE} database \citep{McAlpine2016} and a similar database constructed from the \texttt{ARTEMIS} simulations. These are parameters related to BHs, such as BH masses and mass accretion rates, or to galactic components, such as the gas, stellar and dark matter masses, star formation rates, or galaxy sizes. The identification of bound haloes and subhaloes was performed with {\sc Friends-of-Friends} ({\sc FoF}) \citep{Davis1985} and {\sc{subfind}} algorithms \citep{Springel2001,Dolag2009}. These databases were also used for the identification of galaxy progenitors and the construction of merger trees of MW-mass galaxies.
 
For morpho-kinematical properties of galaxies, we choose to trace the specific stellar angular momentum, $\lambda_{\star}$ (defined as the stellar spin per unit mass), as this is a parameter that is common to all databases and ensures a consistent comparison across all simulations. Moreover, observations highlight $\lambda_{\star}$ as a key property for galaxy evolution, showing close ties to both galactic morphology and disc gravitational instability \citep[e.g.,][]{Romeo2023}. In simulations, $\lambda_{\star}$ correlates well with the disc-to-total mass fraction, $D/T$, as shown by \citet{Thob2019}. In Fig.~\ref{fig:Stars_spin_vs_D_T_Recal} we show the $\lambda_{\star} - D/T$ relation for MW-mass galaxies in the Recal-L025N0752 \texttt{EAGLE} run. Despite a significant scatter, galaxies with disc-like morphologies ($D/T \gtrsim 0.5$) are consistently associated with higher $\lambda_{\star} (\gtrsim 350$~pkpc km s$^{-1}$). 

\section{Results}
\label{sec:Results}

\subsection{The scatter in the BH -- galaxy stellar mass relation}
\label{sec:BH_sr}

\begin{figure*}
\includegraphics[width=2\columnwidth]{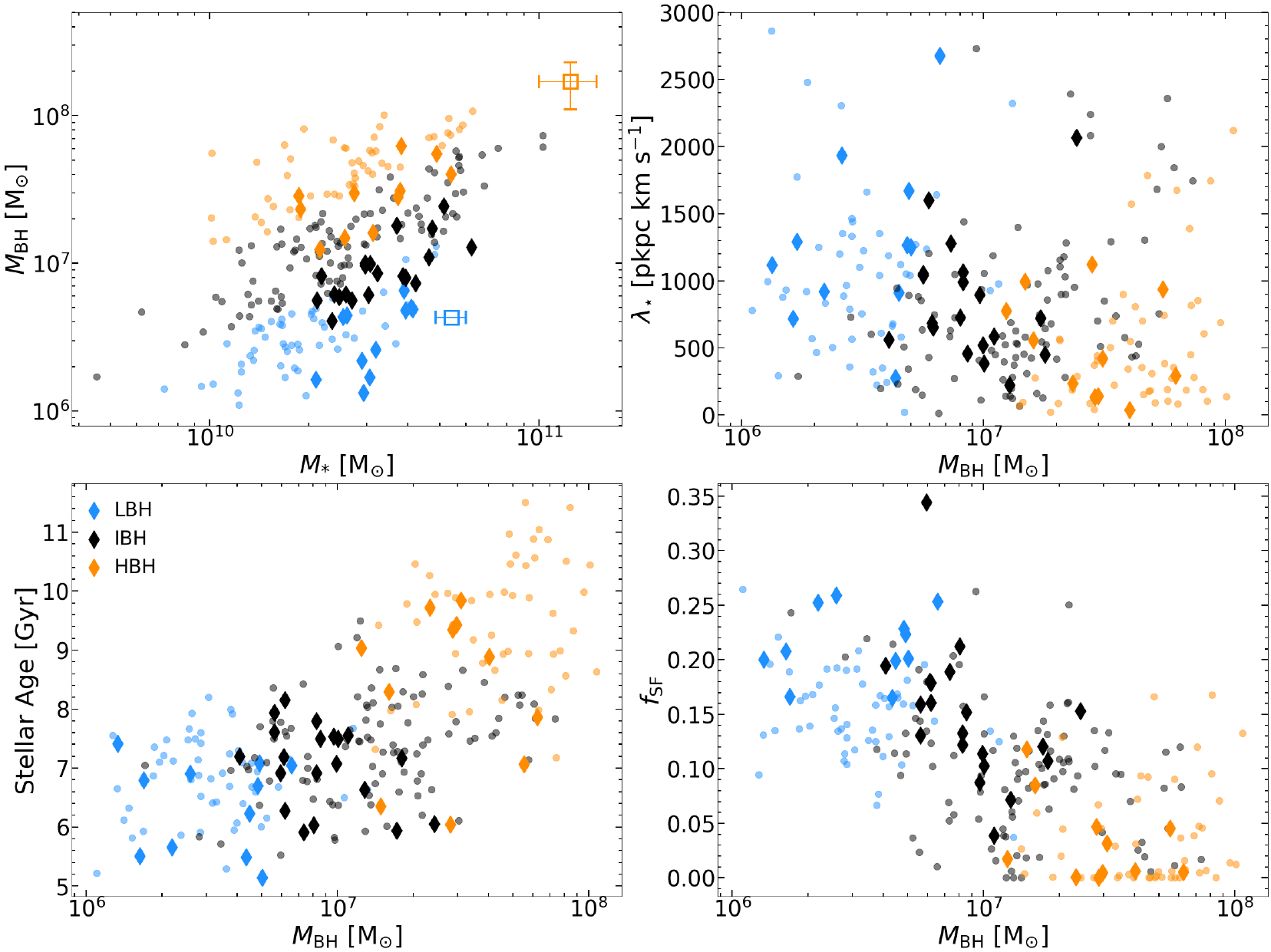}
\caption{Correlations between $M_{\mathrm{BH}}$ and average galaxy properties at $z=0$ in the \texttt{ARTEMIS} (diamonds) and \texttt{EAGLE} (circles) simulations. From top-left to bottom-right, we show the $M_{\mathrm{BH}}$ correlations with $M_\star$, ${\lambda}_\star$, average galaxy stellar age and the fraction of star-forming gas. Systems with low, intermediate and high $f_{\mathrm{BH}}$ values are coloured in blue, black and orange, respectively. The squares with error bars in the top-left panel correspond to data for MW (blue) and M31 (orange). On average, galaxies with higher $f_{\mathrm{BH}}$ tend to have more massive BHs, lower rotational support, older stellar populations, and less gas available for star formation.}
\label{fig:MBH_scaling_relations}
\end{figure*}

Fig.~\ref{fig:Mstar_MBH_relation} shows the $M_{\mathrm{BH}}-M_{\mathrm{*}}$ relation for MW-mass galaxies in simulations versus observations. The \texttt{ARTEMIS} galaxies (left panel) are shown with diamonds and galaxies from  the \texttt{EAGLE} Ref-L050N0752 suite (right panel) are shown with circles. The colours correspond to ${\lambda}_\star$, which serves as a proxy for morphology. Triangles show observational data for early-type (red) and late-type (blue) galaxies compiled by \citet{Greene2020}, while light blue and orange squares represent the measurements for MW and M31. The dashed lines indicate the linear fits derived from these observational data. The solid red (blue) lines in the right panel correspond to linear fits to \texttt{EAGLE} galaxies with $\lambda_{\star}$ below (above) the $350~\mathrm{pkpc~km~s^{-1}}$ threshold. 

The agreement in the normalizations of the $M_{\mathrm{BH}}-M_{\mathrm{\star}}$ relations from simulations with the observations is expected, given the calibration of the BH model \citep{Booth2009}. The good agreement with the slopes, particularly the steeper slope for early-type galaxies compared to late-type galaxies, can be considered to be a success of the models \citep[see also the discussion in][]{Schaye2015}. The simulations also retrieve the observed correlation in the scatter of black hole masses at fixed $M_{\mathrm{\star}}$ with morphological properties, i.e., at fixed $M_\star$, early-type (low $\lambda_{\star}$) galaxies tend to host more massive BHs than late-type systems (high $\lambda_{\star}$).

In particular, MW and M31 represent good examples of scatter in observed BH masses at (nearly) fixed  $M_{\mathrm{\star}}$. They also lie in the vicinity of the two morphological types: MW is clearly associated with late-type galaxies, while M31 lies around the transition zone between late- and early-type galaxies. We note that the observational sample of \citet{Greene2020} includes galaxies with even lower BH masses than that of the MW and higher BH masses than that of M31 (at similar $M_{\star}$ values\footnote{The observational sample of \citet{Greene2020} also includes a wider stellar mass range than selected from the simulations, hence it also shows the evolution of BH masses and galaxy morphological types with stellar mass. Here, however, we are concerned mainly with the scatter in $M_{\mathrm{BH}}$ at nearly fixed $M_{\mathrm{\star}}$.}), which suggests that the models may not explain the full observed scatter. This was also noted by \citet{Roberts2026}, who compared another \texttt{EAGLE} run (the reference model in a $100$~cMpc box) with the observations of \citet{GrahamSahu2023a}.

To further understand the origin of this scatter, we select from the simulations two samples of galaxies with low and high $M_{\mathrm{BH}}/M_{\mathrm{*}}$ ratios (henceforth denoted $f_{\mathrm{BH}}$) to study their evolution. Fig.~\ref{fig:Mstar_MBH_relation} suggests that these two categories should have different morphologies and possibly different formation pathways. 
We therefore classify galaxies in both simulations into three populations according to their $f_{\mathrm{BH}}$ ratio. Galaxies with $f_{\mathrm{BH}}$ above the 75$^{\rm th}$ percentile of their sample are defined as high $f_{\mathrm{BH}}$ (HBH, hereafter), and those with $f_{\mathrm{BH}}$ below the 25$^{\rm th}$ percentile are defined as low $f_{\mathrm{BH}}$ (LBH); the rest of galaxies are considered intermediate $f_{\mathrm{BH}}$ (IBH). 
Fig.~\ref{fig:MBH_scaling_relations} shows various correlations between $M_{\mathrm{BH}}$ and host properties at $z=0$, where the three populations are shown in different shades (orange for HBH, blue for LBH and black/gray for IBH). Data from each simulation are represented with different symbols (diamonds for \texttt{ARTEMIS} and circles for \texttt{EAGLE}).

\begin{figure*}
\includegraphics[width=2\columnwidth]{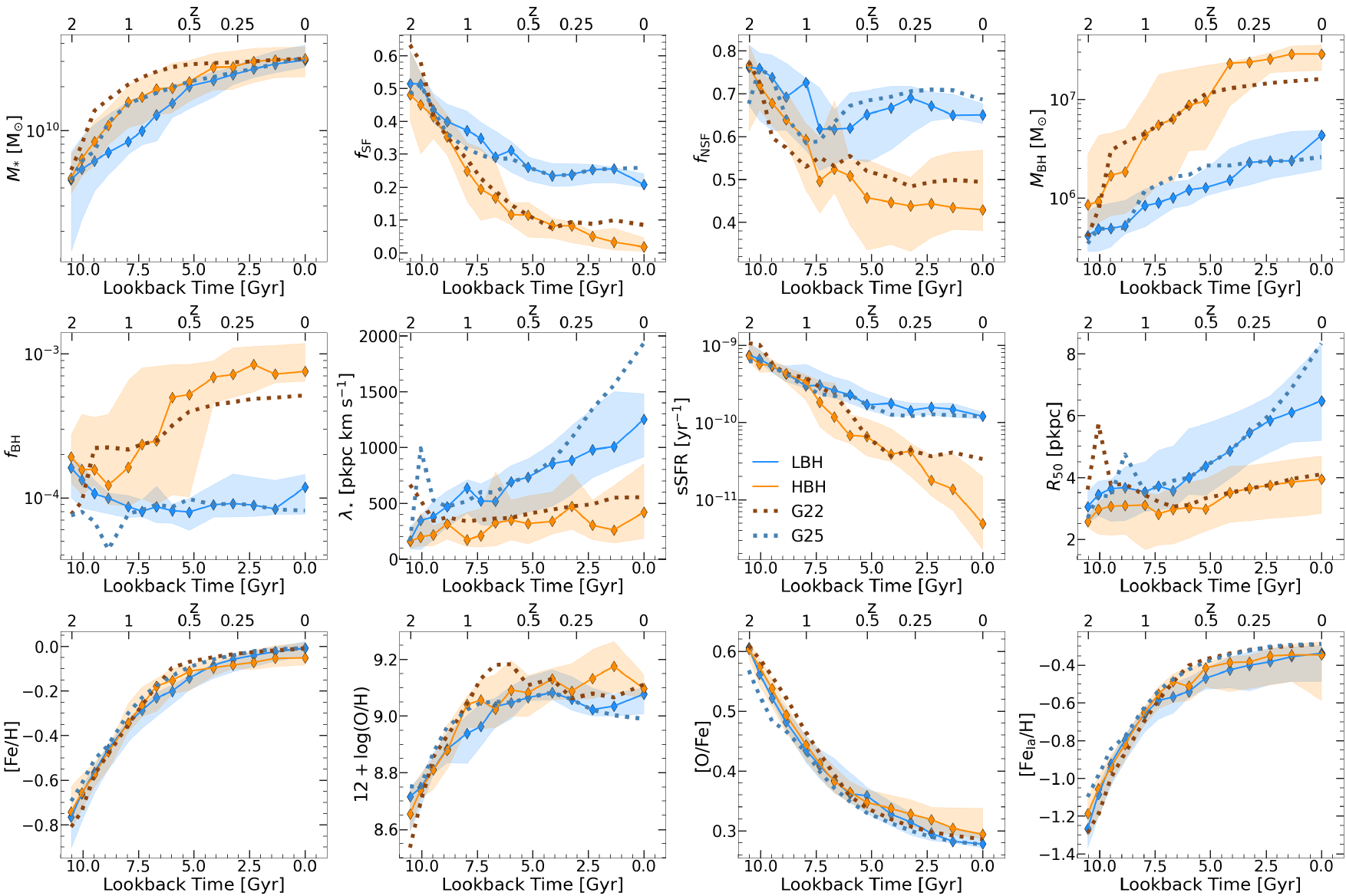}
\caption{Evolution of median properties for both HBH (solid orange line with diamonds) and LBH (solid blue line with diamonds) galaxies in \texttt{ARTEMIS}. The shaded regions enclose the corresponding 25$^{\rm th}$ and 75$^{\rm th}$ percentiles. 
Dotted lines trace the evolution of two individual galaxies in the LBH (G25, blue) and HBH (G22, orange) samples.
First row, in order from left to right show: the stellar mass, star-forming gas fraction, non-star-forming gas fraction, and BH mass. Second row, from left to right, show: the BH-to-stellar-mass ratio, stellar spin, specific star formation rate, and half-mass radius. Third row, from left to right show: the stellar [Fe/H], SF gas (O/H), stellar [O/Fe], and SNIa-based stellar [Fe/H]. With the exception of stellar mass and chemical abundances, all other properties analysed exhibit strikingly different evolutionary trends in LBH vs. HBH galaxies, since at least  $z\approx 2$.}
\label{fig:evo_properties}
\end{figure*}

The top left panel of this figure shows the $M_{\mathrm{BH}}-M_{\mathrm{\star}}$ relation again, this time highlighting the three $f_{\mathrm{BH}}$ populations in each simulation. We note that the $f_{\mathrm{BH}}$ thresholds separating the three populations are slightly higher in \texttt{EAGLE} than in \texttt{ARTEMIS}, indicating that, at fixed $M_\star$, BHs are slightly more massive in \texttt{EAGLE}. This offset may arise from differences in SN feedback parameters adopted in these simulations (see section~\ref{sec:subgrid_models}). Alternatively, it could be due to the larger volume covered by \texttt{EAGLE} ($L=50~{\rm cMpc}$) and possibly sampling denser environments where BH growth through mergers may be more efficient.  Nevertheless, both MW and M31 remain close to their corresponding $f_{\mathrm{BH}}$ categories. 

The top-right panel shows the $\lambda_{\star} - M_{\mathrm{BH}}$ distributions. Although these show a larger scatter than the $M_{\mathrm{BH}}-M_{\mathrm{\star}}$ relation, a clear correlation is seen between BH masses and stellar spins (i.e., morphology):  galaxies with early-type morphologies (low $\lambda_{\star}$) tend to have more massive BHs (high $f_{\mathrm{BH}}$) than late-type ones. Galaxies hosting more massive BHs are expected to be more affected by feedback over time; this leads to quenching of star formation by heating the gas and gas outflows. In section~\ref{sec:Merger_Histories} we will discuss how mergers destabilize the rotational support in galaxies and fuel the growth of BHs; in turn, SN and/or AGN feedback may further inhibit the regrowth of a stable disc.

The bottom-left panel shows the relation between $M_{\mathrm{BH}}$ and the mean stellar ages in simulated galaxies. The ages are computed as $\sum_i \tilde{m}_i(t-\tilde{t}_i)/\sum_i \tilde{m}_i$ \citep{McAlpine2016}, where $t$ is the cosmic time, $\tilde{t}_i$ is the formation time of the star particle $i$ and $\tilde{m}_i$ is the corresponding initial mass. This shows that galaxies with more massive BHs formed their stars earlier.  The bottom-right panel shows the distributions of BH mass as a function of the star-forming (SF) gas fraction in host galaxies.  The SF gas fraction is defined as $f_{\mathrm{SF}} =M_{\mathrm{SF}}/(M_{\star}+M_{\mathrm{SF}})$, where $M_{\mathrm{SF}}$ is the total mass of gas available for star formation. Galaxies hosting more massive BHs tend to have lower $f_{\mathrm{SF}}$, suggesting that SN/AGN feedback has been more effective at heating or removing the gas in these systems.

\begin{figure*}
\includegraphics[width=2\columnwidth]{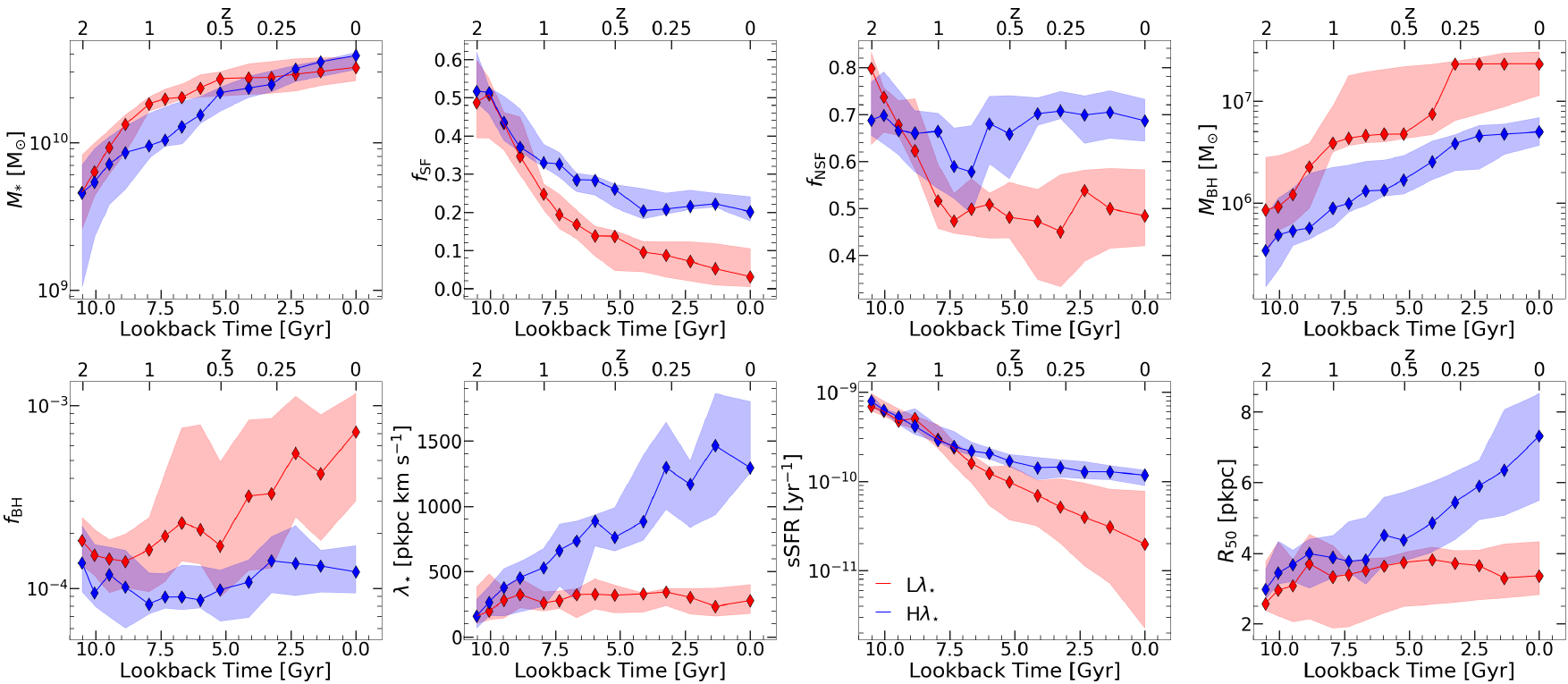}
\caption{Evolution of median properties for both low $\lambda_{\star}$ (solid red line with diamonds) and high $\lambda_{\star}$ (solid  blue line with diamonds) galaxies in \texttt{ARTEMIS}. The shaded regions enclose the corresponding 25$^{\rm th}$ and 75$^{\rm th}$ percentiles. For the nomenclature, see Fig.~\ref{fig:evo_properties}. The differences we obtained previously for the evolution of the LBH and HBH populations are consistent with those obtained for galaxies with high and low $\lambda_{\star}$, respectively.}
\label{fig:evo_properties_V2}
\end{figure*}

\subsection{The co-evolution of BH growth and host properties in galaxies with high and low BH-to-stellar-mass fractions}
\label{sec:Evolution_properties}

Our results show that BH mass correlates with several key galaxy properties at $z=0$. To explore the origin of these correlations, we tracked the main progenitors back in time and studied their evolution. For clarity, we present only the results from the \texttt{ARTEMIS} simulations. 

Fig.~\ref{fig:evo_properties} shows the evolution of LBH and HBH populations introduced before, relating BH growth to the evolution of: stellar masses, the fraction of star-forming gas ($f_{\rm SF}$), the fraction of non-star-forming gas ($f_{\rm NSF} = M_{\rm NSF} / (M_{\star}+M_{\rm NSF})$, where $M_{\rm NSF}$ is the gas mass not available for star formation), stellar spin (${\lambda}_\star$), the specific star formation rate (sSFR), half-mass radius ($R_{50}$), stellar chemical abundances ([Fe/H], [O/Fe] and SNIa-based [Fe/H] = $[{\rm Fe}_{\rm Ia}/{\rm H}]$), and gas-phase metallicity ($12+\log{\rm (O/H)}$).  The solid blue and orange lines show the running medians for the LBH and HBH populations, respectively. The shaded regions enclose the corresponding 25$^{\rm th}$ and 75$^{\rm th}$ percentiles for each parameter. We also highlight two representative galaxies, G25 and G22 (with dotted curves). These galaxies are typical members of the two populations and are investigated in detail in section~\ref{sec:individual_galaxies}.

Although the stellar masses of both categories reach similar values at $z=0$, HBH galaxies tend to grow their stellar mass faster at early times (between redshifts $2-1$), while LBH systems display a more gradual increase, picking up at late times ($z<1$). The enhanced star formation at early times in HBH galaxies is accompanied by a faster growth in BH mass compared to that in LBH systems. The evolution of $f_{\rm BH}$ shows a quick divergence in the two categories from $z\simeq 1$ onward, increasing sharply for HBH galaxies and remaining almost constant, at very low values ($\approx 10^{-4}$), for LBH systems. This indicates that in HBH galaxies BHs grow very efficiently compared with the stellar component, while in LBH galaxies, they both grow at similar rates.  

The evolution of gas fractions also shows significant differences. Both star-forming ($f_{\rm SF}$) and non-star-forming ($f_{\rm NSF}$) fractions start with similarly high values at $z\approx2$, after which they begin to diverge. The LBH population maintains consistently higher gas fractions;   $f_{\rm NSF}$ remains high until $z=0$ ($\simeq 0.6-0.7$), while $f_{\rm SF}$ undergoes only a slow decline. In contrast, in HBH systems, these fractions drop significantly (e.g., $f_{\rm SF}\lesssim 0.1$). This suggests that HBH galaxies experience more efficient gas outflows, due to BH growth. Consequently, the star-forming gas reservoir is depleted, and galaxies quench (most HBH galaxies have sSFR $< 10^{-11}~{\rm yr}^{-1}$ at present). In contrast, the average sSFR in LBH galaxies at $z=0$ is $\simeq 10^{-10}~{\rm yr}^{-1}$, a value which is similar to that of the present-day MW disc. 

The morphological properties of the two populations paint a similar picture: at early times ($z \approx 2$), both types of systems have low rotational support (${\lambda}_\star \approx 200~{\rm pkpc~km~s^{-1}}$); after that, galaxies with low BH masses gradually increase their stellar spin, becoming more disc-like, while those hosting more massive BHs maintain their elliptical shapes; on average, ${\lambda}_\star$ increases by a factor $\approx 6$  in LBH galaxies and only by a factor of $\approx 2$ in HBH systems. The galaxy sizes, $R_{50}$, grow by factors of $\approx 3$ and $\approx 1.6$, respectively. The higher sSFR in LBH galaxies maintains a high rotational support throughout these galaxies, preventing the growth of central BHs, whereas in HBH systems the depletion of the cold gas reservoir through infall onto the BH and the heating/expulsion of gas by outflows lead to quenching of star formation and elliptical morphologies. 

Interestingly, the differences in the chemical evolution between the two populations are much less evident. The evolutions of stellar abundances ([Fe/H], [O/Fe], and $[{\rm Fe}_{\rm Ia}/{\rm H}]$) are almost identical for the LBH and HBH populations, and the scatter in their distributions is very tight (apart from that of $[{\rm Fe}_{\rm Ia}/{\rm H}]$ at lower redshifts). These results are explained by the similar stellar masses in the two galaxy populations and the well-known relation between stellar mass and metallicity that these systems match. We also note that this scaling relation is primarily sensitive to SN feedback, which implies that metal enrichment must have been largely achieved during the early times of enhanced star formation, before most of the growth of central BHs and their possible AGN feedback. 

The differences in the gas-phase metallicity ([O/H]) are also small. The slightly higher values in HBH galaxies at $z < 1$ may reflect their higher $M_\star$ at these epochs, or their more active merger histories (see section~\ref{sec:Merger_Histories}).  In addition, the lower gas mass fractions of HBH galaxies allow stellar feedback to enrich the lower-mass, diffuse gas reservoir more efficiently. Although a higher gas-phase metallicity would normally lead to higher stellar metallicities, as newly formed stars inherit the chemical composition of their natal gas, the much lower sSFRs of HBH galaxies at late times suppress the increase of their stellar metallicities.

\begin{figure*}
\centering
\includegraphics[width=2\columnwidth]{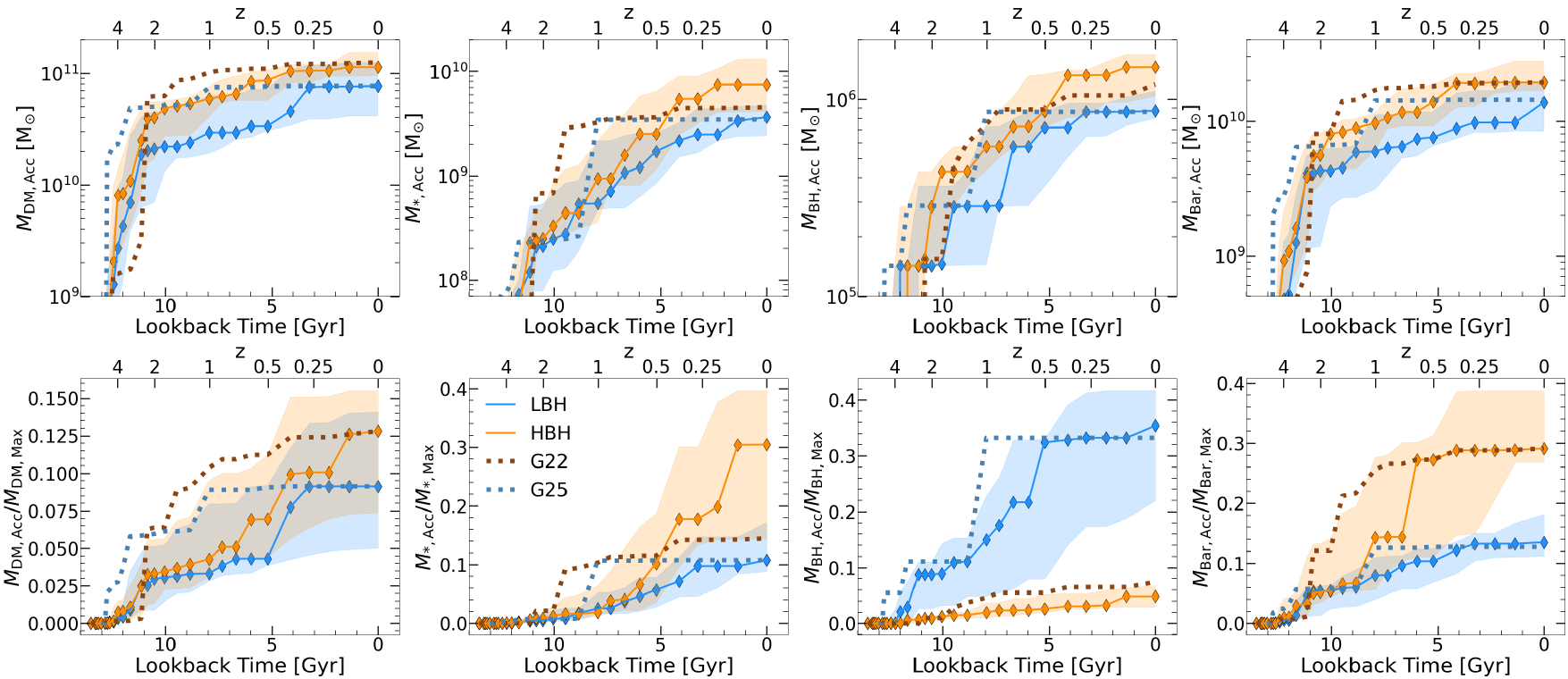}
\caption{Upper panels: cumulative mass accreted via mergers for different galaxy components. Bottom panels: cumulative accreted mass accreted for the same components shown in the upper panels, but normalized by the maximum mass reached in the corresponding component by each galaxy. From left to right: dark matter mass, stellar mass, BH mass, and baryonic mass. Symbols, line styles and colours follow the same convention as in Fig.~\ref{fig:evo_properties}.}
\label{fig:accreted_masses}
\end{figure*}

\subsection{The evolution of galaxies with high and low stellar spin}
\label{sec:Morphology}

Our findings show that the growth of central BHs is closely linked with the morphological changes of host galaxies. As shown in Fig.~\ref{fig:evo_properties}, a clear dichotomy in morphology (traced by low/high $\lambda_{\star}$) exists between galaxies with massive BHs versus those with less massive BHs. However, the $\lambda_\star - M_{\rm BH}$ correlation is not very tight (see Fig.~\ref{fig:MBH_scaling_relations}). It is therefore instructive to examine the evolution of galaxies with high and low $\lambda_\star$ separately. This would show how the properties of galaxies that are disc-like or elliptical today evolve with time, and when these start to diverge. 

We perform a similar analysis as before, but instead separate galaxies as a function of their $\lambda_{\star}$ at $z=0$. For brevity, we refer to systems in the lower 25$^{\rm th}$ percentile ($\lambda_{\star}\approx 474~{\rm pkpc~km~s^{-1}}$) as ${\rm L}\lambda$, those in the upper 75$^{\rm th}$ percentile ($\lambda_{\star}\approx 1105~{\rm pkpc~km~s^{-1}}$) as ${\rm H}\lambda$, and to the intermediate population as ${\rm I}\lambda$. Fig.~\ref{fig:evo_properties_V2} shows the evolution of ${\rm H}\lambda$ and ${\rm L}\lambda$.  Generally, these trends  resemble those of the high- and low-$f_{\rm HB}$ systems, respectively (see Fig.~\ref{fig:evo_properties}). This is due to the significant overlap between the samples selected in terms of their spin or BH mass; specifically, we find that the ${\rm H}\lambda$ sample includes 8 LBH (out of the $11$ systems) and $3$ IBH galaxies (out of $20$), while the ${\rm L}\lambda$ sample includes $7$ HBH (out of the $11$) and $4$ IBH systems.

More elliptical galaxies (${\rm L}\lambda$) tend to form their stellar mass earlier, similar to what was observed for systems with more massive BHs (HBH), and consistent with their older stars (Fig.~\ref{fig:MBH_scaling_relations}). This is indicative of an earlier formation time of the haloes inhabiting these galaxies and an early cessation of star formation. The evolutions of $M_{\rm BH}$ (and $f_{\rm BH}$) in the two $\lambda_\star$ populations start to diverge early, between $z \approx 2-1$ (if not earlier for $M_{\rm BH}$), in agreement with the behaviour displayed by the HBH and LBH populations. However, the separation between these two divergent paths is less pronounced in the case of morphological types than for $f_{\rm BH}$ types (compare Fig.~\ref{fig:evo_properties_V2} with Fig.~\ref{fig:evo_properties}). This indicates that morphologies do not precisely trace the growth of BHs, as seen before. Nevertheless, there is clear evidence that BHs grow faster in elliptical galaxies, particularly in systems that have established this morphology at very early times. These results are similar to those obtained by \citet{Roberts2026} using a sample of galaxies from the \texttt{EAGLE} simulations and using the co-rotational parameter $\kappa_{\mathrm co}$ to distinguish morphological types (although in the case of separating galaxies by $\kappa_{\mathrm co}$ the differences between the two evolutions appear slightly less pronounced than when using $\lambda_{\star}$).

Interestingly, the scatter in the evolutionary paths shown in Fig.~\ref{fig:evo_properties_V2}  differs between the two populations (see also Fig.~\ref{fig:evo_properties}). For example, elliptical galaxies (high BH mass) exhibit a larger scatter in their BH mass growth histories. This may indicate that ellipticals form through a wider range of merger histories, whereas disc-like galaxies require more specific conditions, such as relatively quiescent assembly histories compared to the typical galaxy in this mass range. We investigate these differences in merger histories in section~\ref{sec:Merger_Histories_Averages}.

Both $f_{\mathrm{SF}}$ and $f_{\mathrm{NSF}}$ decrease with time (hence also the sSFR), more strongly so for elliptical galaxies. This is consistent with the behaviour seen for galaxies hosting more massive BHs, in which gas reservoirs were depleted early by more efficient feedback. The evolution of galaxy sizes, $R_{50}$, largely resembles that of $\lambda_{\star}$ (as seen also in Fig.~\ref{fig:evo_properties}). However, differences in galaxy sizes are more clearly seen between samples selected morphologically ({$\lambda_{\star}$) than by $f_{\rm BH}$, which is expected.

\begin{figure*}
\includegraphics[width=2\columnwidth]{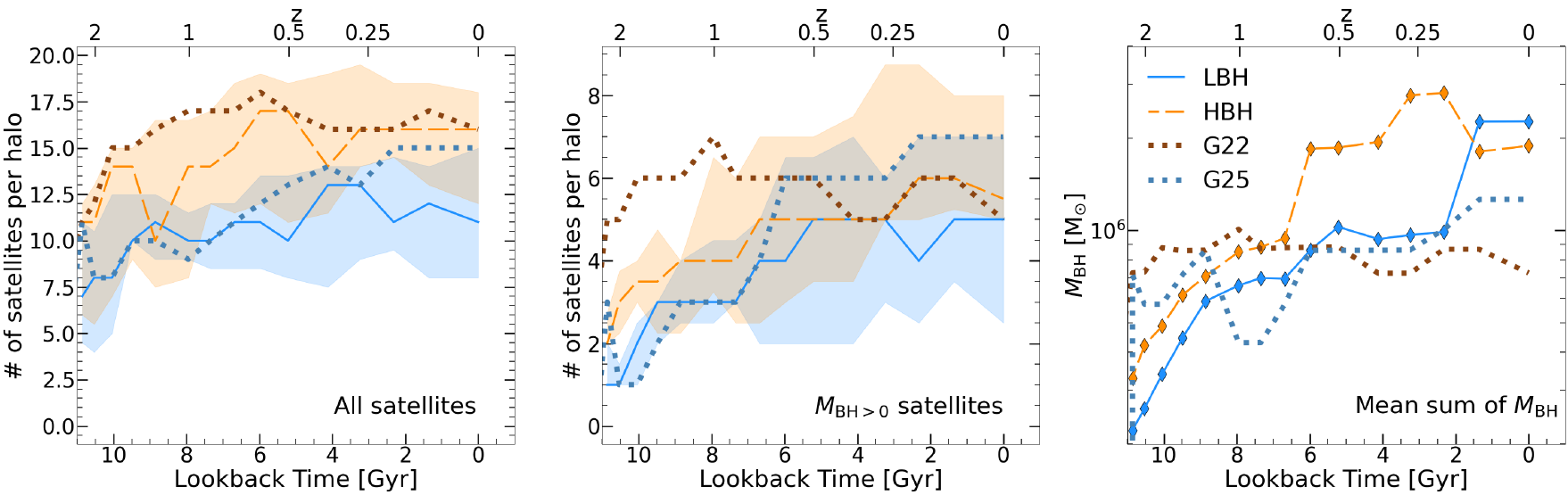}
\caption{Left panel: the median number of satellite galaxies (and scatter) in LBH and HBH hosts across redshift. Only satellites with $M_{\star}\ge 10^{6}~\mathrm{M_{\odot}}$ are considered. Middle panel: the median number of satellites with black holes in the two samples.  Right panel: mean of the total black hole mass in satellites hosting BHs. Symbols, line styles and colours follow the same convention as in Fig.~\ref{fig:evo_properties}.}
\label{fig:n_satellites}
\end{figure*}

\subsection{The role of mergers in the BH growth}
\label{sec:Merger_Histories}

As shown earlier, the central BH masses correlate well with host galaxy properties, including morphologies, in that galaxies with more massive/less massive BHs tend to be elliptical/discy, and vice versa. The different evolutionary paths seem to emerge after $z\approx2$. In the following, we investigate how merger histories can influence these evolutionary paths. For simplicity, we will focus only on the two $f_{\rm BH}$ populations, LBH and HBH.

\subsubsection{The merger-induced growth of BHs and galaxy components}
\label{sec:Merger_Histories_Averages}

Fig.~\ref{fig:accreted_masses} shows the evolution of the mass accreted through mergers in the dark matter, stellar components and central BHs of the two samples. Specifically, at each snapshot, we compute the mass of all subhaloes\footnote{In a few cases where subhaloes are temporarily missed by {\sc{subfind}}, we locate them in the immediate snapshot where they could be identified.} that will merge with the main halo in the following snapshot and sum over their masses to obtain the total accreted mass at that time. We compute the accreted masses for dark matter ($M_{\rm DM,Acc}$), stars ($M_{{\star,}{\rm Acc}}$),  baryons, i.e., gas + stars ($M_{\rm Bar ,Acc}$) and BHs\footnote{As noted before, some subhaloes cannot be identified by {\sc{subfind}} at certain snapshots. This affects the tracking of the BH in these subhaloes also. In this case, we identify the sharp (unphysical) decline in BH mass and insert the value from the immediate snapshot where the subhalo was identified.} ($M_{\rm BH,Acc}$). The plot also includes galaxies G22 and G25 (which will be examined in detail in section~\ref{sec:individual_galaxies}) as representative cases for each sub-sample.

The upper panels of Fig.~\ref{fig:accreted_masses} show the median cumulative  accreted masses for the LBH and HBH populations. $M_{\rm DM,Acc}$ increases sharply before $z \simeq 2$, then more gradually until $z\approx 0.25$ when it flattens. Galaxies with more massive BHs generally accrete more DM, particularly at early times ($z \lesssim 2$). This confirms that the DM haloes hosting galaxies with more massive BHs form earlier.  The accreted baryonic mass shows a trend similar to that of DM, which shows that baryons (mainly the gas brought in by mergers) trace the gravitational potential of DM haloes more accurately.

However, the cumulative BH accreted mass ($M_{\rm BH,Acc}$) increases more gradually, suggesting a delay between the merger-driven supply of gas and its subsequent accretion onto the BH. Interestingly, the accreted mass in the stellar component, $M_{{\star,}{\rm Acc}}$, behaves similarly to $M_{\rm BH,Acc}$, showing a close link between the evolution of the two components. Both $M_{\rm BH,Acc}$ and $M_{{\star,}{\rm Acc}}$ flatten after $z\approx 0.5$, as mergers become less frequent in recent times. 

Overall, galaxies with more massive BHs today (HBH) accreted more mass in all components, which is due to their early formation times and more massive mergers. In particular, the accreted stellar mass in HBH galaxies is a factor of $\approx 2$ higher than for the LBH population by the present time, which is consistent with the similar difference in BH masses between the two populations.

A more informative measure of merger-driven growth is the fraction of accreted mass, i.e., the ratio of the cumulative accreted mass by the corresponding maximum mass of that given component during the whole galaxy evolution. This adjusts for any changes in baryonic mass (e.g., due to gas depletion or stellar evolution) and accounts for intrinsic variations in component masses between galaxies. These results are shown in the bottom panels of Fig.~\ref{fig:accreted_masses}, where we plot the accreted fractions for dark matter ($f_{\rm DM,Acc}=M_{\rm DM,Acc}/M_{\rm DM,Max}$), stars ($f_{{\star ,}{\rm Acc}}=M_{{\star ,}{\rm Acc}}/M_{{\star ,}{\rm Max}}$), BHs ($f_{\rm BH,Acc} =M_{\rm BH,Acc}/M_{\rm BH,Max}$) and baryons ($f_{\rm Bar ,Acc}=M_{\rm Bar ,Acc}/M_{\rm Bar,Max}$). More drastic separations between the evolutions of the two populations can be seen in this case, and the rates of growth are also changed (i.e., indicating more recent accretion times).  

The accreted dark matter mass fraction, $f_{\rm DM,Acc}$, exhibits similar trends in both populations, with HBH galaxies showing slightly larger merger contribution than LBH galaxies.  HBH galaxies have  significantly higher accreted stellar and baryonic mass fractions. By $z=0$, they accumulated $\approx 30\%$ of stellar/baryonic mass through mergers, compared with only $\approx 10-15\%$ for LBH galaxies. As noted before, HBH galaxies also display a wider scatter about the median (see Figs.~\ref{fig:evo_properties} and \ref{fig:evo_properties_V2}), which is even more pronounced for accreted mass fractions ($f_{{\star ,}{\rm Acc}}$ and $f_{{\rm Bar ,}{\rm Acc}}$). This increased scatter reflects the more diverse evolutionary pathways for high BH-mass systems.

The importance of mergers in HBH galaxy evolution is evidenced by the lack of significant rotational support in their stars over time.
Interestingly, the BH mass added by mergers shows different trends from those obtained for the dark matter, gas, and stellar components. In the case of the LBH sample, $\approx 35\%$ of the BH mass at $z\approx 0$ was accreted at $z\gtrsim 0.5$. Fig.~\ref{fig:evo_properties} shows that the corresponding accreted BHs are not so massive ($\lesssim 10^6 ~{\rm M}_{\odot}$). However, the contribution of mergers to the assembly of the BH mass is $\lesssim 10\%$ for the whole sample of HBH galaxies at $z=0$.  These results imply that the more efficient growth of BHs in HBH galaxies is fueled primarily by gas accretion.  

The average accretion patterns analysed so far indicate that the faster growth of BHs in HBH galaxies is mostly due to a more efficient gas accretion onto central BHs and, to a lesser extent, to BH--BH mergers (Fig.~\ref{fig:accreted_masses}). The enhanced gas accretion rate onto BHs in HBH systems may lead to AGN activity and consequent heating of the ISM and gas outflows. This would explain the lower gas fractions and sSFRs obtained for the HBH population, compared to the LBH sample (Fig.~\ref{fig:evo_properties}). Additionally, mergers induce star formation and feedback from SNe, which further lower the gas fractions. The merger-driven assembly hypothesis for HBH galaxies is also supported by their weaker rotational support. Although mergers subside below $z\approx 0.5$ in these systems, as seen from the decrease in the accreted mass, discs cannot form because the gas reservoir is already depleted by this epoch.

\begin{figure*}
\includegraphics[width=2\columnwidth]{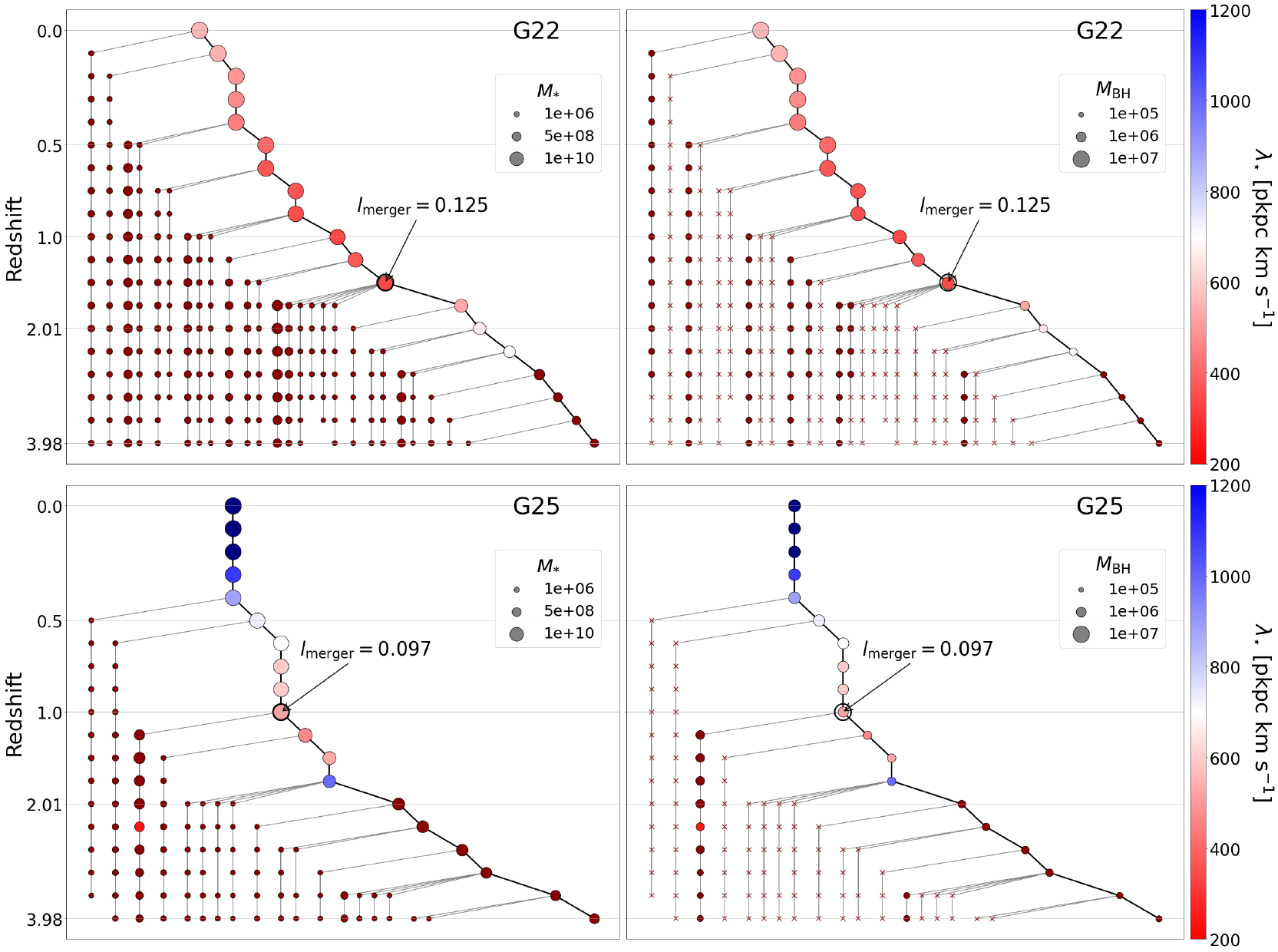}
\caption{Merger trees of galaxies G22 and G25. Each symbol represents a progenitor at a given $z$ (left axes) of the galaxy at the top of the tree. Each gray line along a tree branch connects a progenitor to its immediate descendant at lower $z$. The main branch (i.e., the main progenitor of the respective galaxy) is shown on the right, highlighting the major mergers. Symbols are colour-coded according to $\lambda_{\star}$, with sizes that scale with $M_{\star}$ (left panels) and $M_{\mathrm{BH}}$ (right panels). In the right panels, galaxies without BH are shown with crosses.}
\label{fig:MTs}
\end{figure*}

\begin{figure*}
\includegraphics[width=1.7\columnwidth]{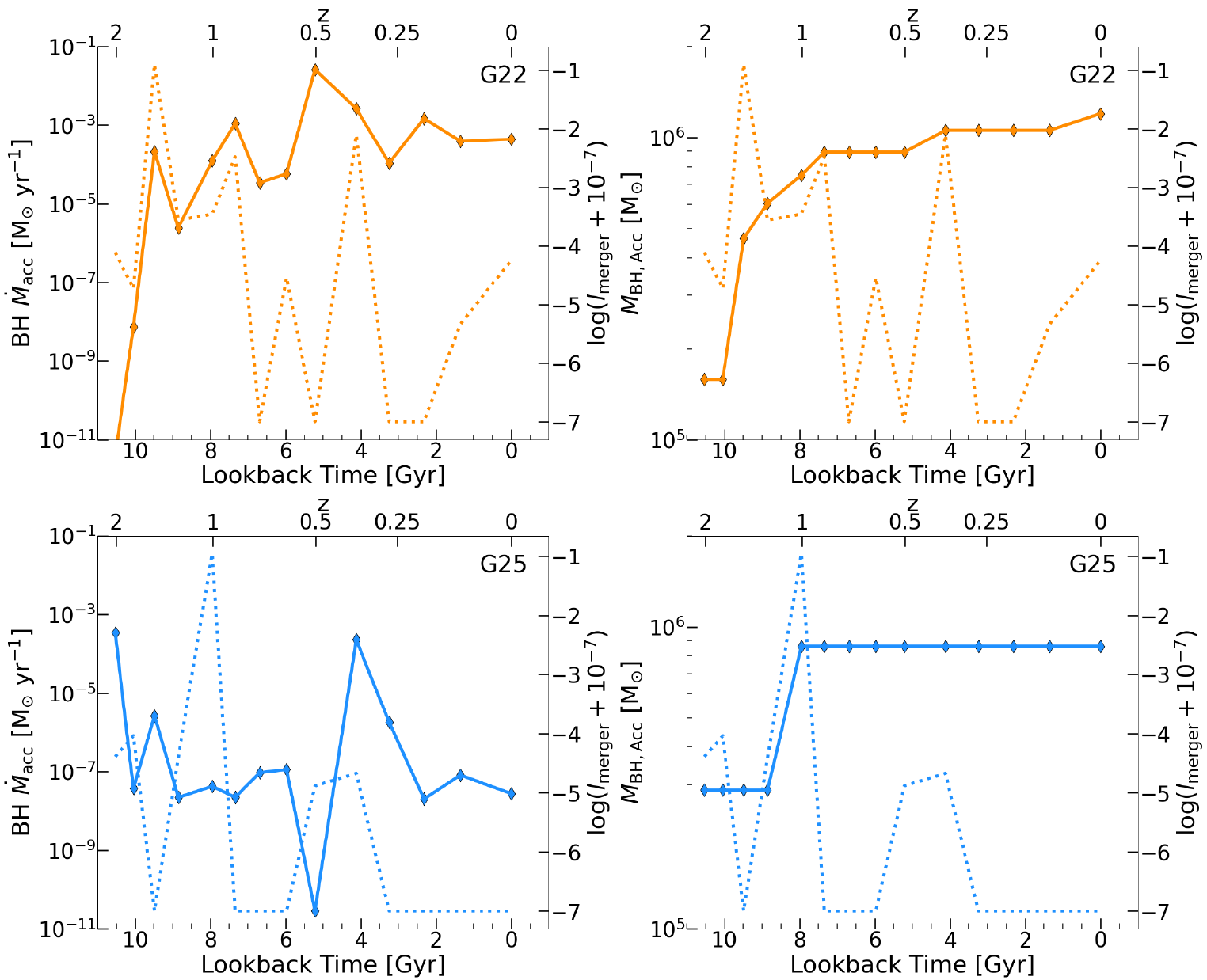}
\caption{The evolution of gas accretion rate onto BHs (left panels, left vertical axes, solid lines) and the growth of BH mass via accretion through BH mergers (right panels, left vertical axes, solid lines) for galaxies G22 (upper panels) and G25 (bottom panels). Also shown is the evolution 
of $l_{\mathrm{merger}}$ parameter (all panels, right vertical axes, dotted lines). Mergers correspond to peaks in $l_{\mathrm{merger}}$. Massive mergers are generally associated with an increase in the gas accretion rate onto BH and/or an increase in the BH accreted mass, though BH--BH coalescence (see text).}
\label{fig:BH_acc_lmerger}
\end{figure*}

\subsubsection{Satellite populations of systems with high and low BH masses}
\label{sec:sats}

So far we examined the global accreted mass. A complementary view on accretion histories is given by the population of dwarf galaxies around their hosts which gives insight into the  contribution of satellites to the accreted host mass at any given time. In addition, this could give us a glimpse into the properties of BHs in satellites (note, however, that the BH seed mass is comparable with the stellar mass of the smaller dwarfs resolved by \texttt{ARTEMIS}, so BHs can be resolved only for a small fraction of satellites). In the following, we track only satellites with $M_{\star}\ge 10^{6}~\mathrm{M_{\odot}}$, to avoid numerical resolution effects. 

The left panel of Fig.~\ref{fig:n_satellites} shows the median number of all satellite galaxies located within the virial radius of their LBH or HBH hosts at redshifts from $z\approx 2$ until the present. There are some notable differences between the two satellite populations, likely caused by the differences in the virial masses of their hosts, namely HBH galaxies being typically more massive than LBH galaxies. Specifically, HBH galaxies have consistently more satellites (ranging between $\sim 11-16$) than LHB galaxies ($7-13$).  At $z\approx 0$, the median numbers are $16$ for HBH and $11$ for LBH systems, respectively. The implication is that galaxies hosting more massive BHs had more active accretion histories, which brought in more gas fuel. These results are also consistent with the higher accreted masses and fractions examined earlier.

The middle panel of Fig.~\ref{fig:n_satellites} includes a similar analysis, but only for satellites that host a BH at the time of accretion.  This shows a faster increase in the number of satellites between $z=2-0.5$, in both samples. This is likely due to the time it takes for satellites to become more massive and therefore host a BH. The median number also increases in both samples, from $1-2$ at $z\approx 2$, to $\sim 6-7$ in HBH galaxies and $\sim 5$ for LBHs. Although these differences are not large in absolute terms, the key difference is in the number of high mass satellites, i.e., those that may result in major mergers (Note that these are satellites that are not yet disrupted, so the results are only indicative of the possibility of mergers). 

In the right panel, we show the mean total BH mass (computed only for satellites hosting BHs) in the two host-galaxy samples. At high redshift ($z\gtrsim 1$), BHs are still low-mass, close to the seed value, and the total BH mass in satellites is $\lesssim 10^{6}\, {\rm M}_{\odot}$. At lower redshift, this value  increases to a few $10^6\, {\rm M}_{\odot}$ in the HBH hosts, $\approx 2$ times higher than the corresponding value in the LBH systems. This reflects the higher number of satellites at any given time in HBH systems, particularly of the high-mass satellites that host more massive BHs. Some of these satellites eventually merge with their host. Below $z\approx 0.25$, the trend is reversed; this is likely due to HBH systems already experiencing their major mergers, whereas LBH galaxies still have high mass satellites orbiting around them. These results suggest that the differences at the high-mass end of the satellite populations, in terms of number of massive satellites and their redshift of accretion, have direct consequences for the growth of central BHs in host galaxies.

\subsubsection{Two case studies for LBH and HBH galaxies}
\label{sec:individual_galaxies}

So far, we analysed the global accretion histories of the two galaxy sub-samples.  We now proceed to explore individual merger histories of two galaxies, each chosen to be typical cases for their parent populations: G25 for LBH galaxies and G22 for the HBHs (see the blue and orange dotted lines in  Figs.~\ref{fig:evo_properties}, \ref{fig:accreted_masses} and \ref{fig:n_satellites}).

Fig.~\ref{fig:MTs} shows the merger trees of these two galaxies, upper panels for G22 and bottom panels for G25. On the left, we show the stellar masses of progenitors, and on the right, the corresponding black hole masses. Symbols scale with the corresponding masses and are colour-coded according to the morphology ($\lambda_{\star}$) of the progenitor at that time. For clarity, we only show progenitors with $M_\star \ge 10^{5}~{\rm M}_\odot$. Progenitors without BHs are shown with crosses. G22 experiences more mergers, of both low- and high-mass (including more minor mergers that occur between consecutive simulation snapshots). Consequently, there are more BHs contributed by mergers in this galaxy than in G25. In addition, G22 has maintained a low $\lambda_{\star}$ (early-type morphology) throughout its evolution. G25 formed a disc early on ($z\approx 2$) and experienced a massive merger at $z\approx 1$ that temporarily changed its morphology to elliptical.  After a final merger, at $z\approx 0.5$, G25 could form a stable disc ($\lambda_{\star}=2000~\mathrm{pkpc~km~s^{-1}}$, see Fig.~\ref{fig:evo_properties}).

To further evaluate the effect of mergers on the host galaxy at any given snapshot, we introduce a parameter called the `level of merger', $l_{\mathrm{merger}}$. This is defined as $\sum_i M_{\star,i}/M_{\star,{\rm MP}}$, i.e., the sum over the stellar masses ($M_{\star,i}$) of all progenitors $i$ outside the main branch at a snapshot prior to when they merge with the main branch, divided by the stellar mass of the main at that snapshot, $M_{\star,{\rm MP}}$. Defined this way,  $l_{\mathrm{merger}}$ enables us to identify the most massive mergers in a system. Fig.~\ref{fig:MTs} shows this parameter for the most massive mergers in G22 and G25. This indicates that G22 experienced a more massive merger than G25, which occurred at $z\approx 1.5$. This event, together with more frequent minor mergers, prevented this galaxy from forming a stable disc.

In Fig.~\ref{fig:BH_acc_lmerger} we investigate the evolution of the gas accretion rate  onto the central black hole (BH ${\dot{M}}_{\rm acc}$) and that of the accreted BH mass ($M_{\rm BH,Acc}$) in the two representative galaxies. We note that, due to the coarse time sampling of the simulation outputs, the accretion rate onto the BH does not accurately capture the instantaneous accretion rate, showing high time variability (this is a well-known effect, discussed in \citealt{McAlpine2016}).  Hence, we only consider the overall trends for this quantity. The gas accretion rate in G22 (top left panel) shows a rapid increase around $z \approx 2$, stabilizing at a value $\approx 10^{-3}~{\rm M}_{\odot}{\rm yr}^{-1}$ until the present, although with some minor fluctuations. The rapid increase phase coincides with the highest peak in $l_{\mathrm{merger}}$, which corresponds to the major merger event experienced by this galaxy at $t_{\mathrm{L}}\approx 10.5~\mathrm{Gyr}$. Through this merger, G22 accreted around $12\%$ of its stellar mass at that time. Several other smaller peaks in $l_{\mathrm{merger}}$ ($< 0.01$) are also visible; these are associated with minor mergers or possibly with fluctuations in the gas accretion rate. The major merger in G22 is also associated with a rapid increase in accreted BH mass, with a corresponding $M_{\rm BH}$ increase of a factor of $\approx 2$. Two minor mergers contribute marginally to $M_{\rm BH,Acc}$, at $t_{\mathrm{L}}\approx 8.5~\mathrm{Gyr}$ and $t_{\mathrm{L}}\approx 4~\mathrm{Gyr}$. Overall, the results show that the BH in G22 grew primarily as a result of an early major event. This event led to more gas accretion onto the main BH, and to the coalescence of the BH of the satellite with the BH of the host.

These results are consistent with the behaviour seen for G22 in Fig.~\ref{fig:accreted_masses}, which indicates that most of the mass accretion occured at $z \gtrsim 1$. The merger tree (Fig.~\ref{fig:MTs}) shows many minor mergers at recent times, with a final such event at $z\approx 0$, however these do not contribute significant mass.  Several of these events contributed  BHs, but of very small mass. The most massive merger ``event'' at $t_{\mathrm{L}}\approx 10.5~\mathrm{Gyr}$ was comprised of 6 individual progenitors that merged onto the main branch at that epoch, only two of them having a BH.

In G25, the typical LBH galaxy, the gas accretion rate (bottom left panel) decreases since $z \approx 2$  to a value of $\approx 10^{-7}~{\rm M}_{\odot}{\rm yr}^{-1}$ at present time, which is about 4 orders of magnitude below the average value for G22 (Fig.~\ref{fig:BH_acc_lmerger}). The most massive merger in this system, at $z\simeq 1$, did not result in an increase in the gas accretion rate but led to a significant increase in the mass acquired through BH--BH coalescence (bottom right panel). An abrupt change occurs in the gas accretion rate later on, at $z\approx 0.5$, which could be caused by the last merger in this system (see Fig.~\ref{fig:MTs} and the change in galaxy morphology at that time). At the time of the most massive merger, at $t_{\mathrm{L}}\approx 8~\mathrm{Gyr}$, G25 accreted around $10\%$ of its stellar mass, while the BH mass increased by a factor of $\approx 3$. No other mergers appear to be significant afterwards (see also  Figs.~\ref{fig:accreted_masses} and ~\ref{fig:MTs}).  

The trends obtained for G22 and G25 are representative for their respective parent populations. HBH galaxies tend to accrete more mass via mergers than LBH systems and experience more minor mergers along their whole formation histories. These richer merger histories lead to an increase of  $M_{\rm BH}$ in HBH systems. This occurs mainly through dissipation of the angular momentum of baryons during mergers, which facilitates the BH growth via gas accretion; in HBH systems BH also grow via direct BH--BH mergers, although this channel contributes less to the total BH mass.

\section{Discussion}
\label{sec:discussion}

\subsection{Comparison with previous studies}
Our study has shown that the scatter in the ${\rm M}_{\mathrm {BH}} -{\rm M}_{\star}$ relation is associated with intrinsic galaxy processes, largely driven by mergers. This is reflected in the distributions of galaxy properties, such as gas fractions, specific star formation rates, galaxy sizes, and stellar spins (morphology), which are correlated with the central BH mass. We have shown that these properties are comparable at $z\approx 2$, which implies that the scatter in the ${\rm M}_{\mathrm {BH}} -{\rm M}_{\star}$ relation of MW-mass progenitors should be much smaller at high redshift. From $z\approx 2$ onward, these galaxies exhibit clear differences in their evolutionary paths. The two samples mainly analysed here, chosen based on their current $f_{\rm BH}$ ($\approx$ BH mass), exhibit strongly divergent paths until $z \approx 0$ in terms of their morphological properties, gas fractions and star formation/quenching state.
These results are not unexpected, as many other studies, using both simulations and observations, have found that black hole mass is a strong predictor of galaxy properties, including their quenching (e.g., \citealt{Piotrowska2022}; \citealt{Bluck2023}; \citealt{Goubert2024}; \citealt{Lim2025}). The approach in our study was to assemble a statistically representative sample of MW-mass galaxies to investigate the connection between BH growth and host-galaxy properties. 

A similar approach was taken recently by \citet{Roberts2026}, who used \texttt{EAGLE} to show that the scatter in the $M_{\rm BH}-M_{200}$ relation correlates with morphology, specifically with the kinematical parameter $\kappa_{\mathrm co}$. They also found that this scatter correlates even more strongly with the halo binding energy, $E_{\rm bind}$, and that mergers (traced by the global accreted fraction in these galaxies) are important in determining the morphological evolution of the main galaxies. The key role of the binding energy in the scatter of $M_{\rm BH}$ at fixed halo mass was also highlighted by \citet{Davies2019}, who linked it to the ability of $L^{\star}$ haloes to expel gas through AGN feedback. The role of mergers in the scatter of the $M_{\rm BH}-M_{200}$ relation in the  $L^{\star}$ regime was also discussed by \citet{Davies2024} (see also \citealt{Davies2022}) who used controlled simulations of a MW-mass galaxy.

Our results suggest that MW-mass galaxies may be affected by multiple mergers and subsequent SN / AGN feedback during their lifetime, and that it is the cumulative effect of these processes that determines their  evolutionary paths. The interplay between these processes is, of course, complex and may not be easily disentangled. At this galaxy mass scale, perhaps SN feedback is still dominant, although AGNs could become important for some galaxies, particularly for those with more massive black holes.  In Appendix~\ref{sec:appendix}, we examine the effects of varying the SN and AGN feedback models in the \texttt{EAGLE} simulations. Stronger SN feedback delays star formation, suppresses BH growth, and leads to higher gas fractions and sSFRs at $z=0$. In contrast, variations in the AGN feedback have little effect on the global galaxy or BH properties, indicating that AGN feedback is not the primary mechanism on these scales. However, AGN remains an important factor in galaxy models, even on these scales. As shown in Fig.~\ref{fig:evo_properties_EAGLE_diff_AGN}, simulations without AGN feedback fail to produce quenched galaxies (sSFR $<10^{-11}~{\rm yr}^{-1}$), demonstrating that this type of feedback is responsible for quenching the population with high BH masses.  

Other cosmological simulations indicate that AGN feedback can affect the morphological properties of galaxies, including those of MW-mass (\citealt{Torrey2020}; \citealt{Getachew-Woreta2022}; \citealt{Irodotou2022}; \citealt{Getachew-Woreta2025}; \citealt{Sivasankaran2025}). For example, using the \texttt{Horizon-AGN} simulation, \cite{Dubois2016} found that AGN feedback can lead to more spherical galaxy shapes, including around the MW-mass range (although MW is near the resolution limit for these simulations).

Our analysis also indicates that BH--BH mergers typically contribute more to the BH mass in LBH galaxies in comparison to those in HBH systems. This result agrees with that of \cite{Bravo2025}, who found that, at fixed galaxy stellar mass, BH growth in galaxies with less massive BHs is merger driven, while in galaxies with more massive BHs it occurs mainly via gas accretion. However, we note that the details of BH modeling, particularly the seeding and dynamics of BHs, can affect how much mergers and gas accretion contribute to their mass assembly \citep[e.g.][]{Pacucci2017, Tremmel2017, Habouzit2021}.

\subsection{Implications for MW and M31}

Based on our classification scheme, MW could be identified as an LBH galaxy, whereas M31 largely falls into the HBH category. Our results indicate that this distinction is physically motivated, arising primarily from differences in their merger histories. The MW has experienced a major merger, Gaia-Enceladus/Sausage (GES), about $8-9~\mathrm{Gyr}$ ago (\citealt{Belokurov2018}; \citealt{Helmi2018}), after which it had an almost quiescent history (e.g., \citealt{Wyse2001,Ruchti2015}). In contrast, M31 probably had a more active merger history, as demonstrated by its thicker and more disturbed stellar disc, the more prominent inner spheroid, and its more numerous satellites and tidal streams (e.g., \citealt{McConnachie2009,McConnachie2018}). The last accretion and disruption of a massive satellite galaxy occurred $\approx 2-3~\mathrm{Gyr}$ ago (e.g., \citealt{Hammer2018}). This also appears to have triggered a burst of star formation \citep{Williams2015,DSouza2018}. Similar differences were noted between the HBH and LBH samples: HBH galaxies have more active merger histories and less rotational support as a result than LBH systems.

Although the simulations do not include exact matches of MW and M31, they suggest that the differences in the BH masses of these two systems could be explained by differences in merger histories. Taking the example of G25, a disc galaxy with low BH mass like the MW, its merger history was fairly quiescent, including a single major merger at $t_L \simeq 8$~Gyr that brought in the gas fuel to catalyse the (re-)formation of the stellar disc. This is similar to the merger of GES with the MW $\simeq 8-9$~Gyr ago, an event that shaped the formation of the Galaxy disc \citep{Belokurov2022}.  In G25, the major merger also contributed $\approx 40$\% of the final BH mass through the coalescence of its BH (with a mass of a few $10^5~{\rm M}_\odot$) with that of the main halo (see Fig.~\ref{fig:MTs} and the blue dotted line in Fig.~\ref{fig:accreted_masses}). In the case of the MW, models of a merger of a central BH from GES with the supermassive black hole present at the center of the MW at that time (a 1:4 mass ratio) can explain the properties of Sgr A$^*$, such as the misalignment of its spin and its high amplitude \citep{Wang2024}. 

In the case of G22, which is a typical example of the HBH population, a substantial fraction of its BH mass was achieved through the enhanced gas accretion rate triggered by at least one major merger (see Fig.~\ref{fig:BH_acc_lmerger}). Similarly, the more massive merger in M31 than in MW could have driven gas inflows more efficiently, feeding the BH. It is also plausible that, given its current high BH mass, M31 could have experienced some AGN activity in the past. Although its BH appears to be currently dormant, there are some indications that brief AGN episodes occurred recently (e.g., \citealt{Zhang2019}).

\section{Conclusions}
\label{sec:concl}

We investigated the origin of the scatter in $M_{\rm BH}$ at fixed $M_\star$ using samples of MW-mass galaxies from the \texttt{EAGLE} simulations and the higher resolution, zoomed-in \texttt{ARTEMIS} simulations. 
We retrieved previously reported correlations between $M_{\rm BH}$ (at fixed $M_\star$) and stellar spin, mean stellar age, and the star-forming gas fraction, reflecting the coupled evolution of these galaxy properties. Our simulations indicate that the scatter in BH masses among $L^{\star}$ galaxies is driven primarily by differences in their assembly histories, that start to diverge around $z\simeq 2$. Galaxies hosting lower-mass BHs experience relatively quiescent merger histories, retain more disc-dominated morphologies, and grow their BHs through comparable contributions from gas accretion and BH--BH mergers. By contrast, galaxies hosting more massive BHs undergo more active merger histories and tend to become elliptical. These mergers increase $M_{\rm BH}$ primarily by triggering enhanced gas accretion, a channel that accounts for the majority of the final BH mass, and through BH--BH mergers. The resulting AGN feedback is sufficient to quench this subset of high BH-mass $L^{\star}$ galaxies, despite the overall $L^{\star}$ population remaining primarily regulated by SN feedback. Overall, these trends are broadly consistent with observations of the Milky Way and M31, two galaxies at the other end of the BH-mass spectrum in the $L^{\star}$ regime.

The conclusions of the main plots can be summarised as follows:

\begin{itemize}
    \item  At fixed $M_\star$, $M_{\rm BH}$ correlates with morphology, consistent with observations \citep[e.g.,][]{Greene2020}. Galaxies hosting less massive BHs are predominantly disc-dominated (high stellar spin), whereas those with more massive BHs are more spheroidal (Fig.~\ref{fig:Mstar_MBH_relation}). Galaxies with more massive BHs form their stars earlier and contain less star-forming gas owing to cumulative feedback (Fig.~\ref{fig:MBH_scaling_relations}).
    The evolution of galaxies with the lowest and highest $f_{\rm BH}$ (or $M_{\rm BH}$) diverges after $z\approx 2$. By $z=0$, an order-of-magnitude difference in $M_{\rm BH}$ corresponds to at least a factor-of-two difference in sSFR, gas fraction, and galaxy size(Fig.~\ref{fig:evo_properties}), indicating that $M_{\rm BH}$ is a stronger predictor of gaseous galaxy properties than morphology (Fig.~\ref{fig:evo_properties_V2}). 
    
    \item High-$M_{\rm BH}$ galaxies accrete more dark matter, stars, gas, and BH mass than their low-$M_{\rm BH}$ counterparts, consistent with their larger halo masses (Fig.~\ref{fig:accreted_masses}). Normalized by the final mass of each component, mergers contribute a larger fraction of the stellar mass ($\sim30\%$ in high-$M_{\rm BH}$ galaxies versus $\sim10\%$ in low-$M_{\rm BH}$ galaxies), and a smaller fraction of the dark matter mass in both. In contrast, BH--BH mergers contribute a larger fraction of the final BH mass in low-$M_{\rm BH}$ galaxies ($\sim40\%$) than in high-$M_{\rm BH}$ systems ($\lesssim10\%$). Consequently, gas accretion dominates BH growth in high-$M_{\rm BH}$ galaxies ($\sim90\%$), whereas gas accretion and BH--BH mergers make more comparable contributions in low-$M_{\rm BH}$ galaxies. M31 and MW  may exemplify these two BH growth channels. This scenario is plausible given their different merger histories.

    \item  High-$M_{\rm BH}$ galaxies experience more active merger histories, which reduce rotational support, drive gas inflows, and enhance BH growth (Fig.~\ref{fig:accreted_masses}). Although merger activity declines after $z \approx 0.5$, these systems cannot (re)form stellar discs. High--$M_{\rm BH}$ galaxies have more satellites at any given time, including more high-mass satellites that host their own BHs (Fig.~\ref{fig:n_satellites}). Massive satellites contribute to the growth of the central BH in the host galaxy by inducing gravitational instabilities that trigger star formation and via direct BH--BH mergers.
    
    \item The contrasting merger histories of two representative galaxies,  G22 and G25, illustrate these trends (Figs.~\ref{fig:MTs} and \ref{fig:BH_acc_lmerger}). G22 experienced more minor mergers, continuing to $z\approx 0$, and a  major merger at $z \approx 1.5$, preventing the formation of a stable disc. By contrast, G25 formed a disc at $z\approx 2$ and reformed it after a major merger at $z\approx 1$, remaining quiescent thereafter (Fig.~\ref{fig:MTs}). The more active merger history of G22 also drove stronger gas inflows, enhancing BH gas accretion by factors of a few relative to G25 (Fig.~\ref{fig:BH_acc_lmerger}). 
    
   \item Feedback also plays a key role in regulating BH growth and galaxy evolution. Varying the strengths of SN and AGN feedback independently shows that SN feedback is the primary regulator of MW-mass galaxies (Fig.~\ref{fig:evo_properties_EAGLE_diff_FB}), whereas AGN feedback is essential for quenching them, becoming increasingly effective in systems hosting more massive BHs (Fig.~\ref{fig:evo_properties_EAGLE_diff_AGN}).

\end{itemize}

Together, these results suggest that the diversity in the assembly histories of $L^{\star}$ galaxies naturally gives rise to the observed scatter in BH masses. In addition to shaping the BH growth, differences in accretion history also affect the host galaxy morphology and the effectiveness of AGN feedback. At the $L^{\star}$ scale, the AGN feedback becomes important only in the most massive BH systems, while the SN feedback continues to regulate the galaxy population as a whole.

 \section*{Acknowledgements}
The authors thank the reviewer for constructive feedback that helped improve the clarity of the paper. The authors also thank the \texttt{EAGLE} team for making their  code available for the \texttt{ARTEMIS} project. Various Python libraries have been used in this study, including \textsc{h5py} (\url{http://www.h5py.org/}) and \textsc{AstroPy} \citep{Astropy2022}, as well as the publicly available \textsc{read\_eagle} module (\url{https://github.com/jchelly/read_eagle}, \citealp{eagle2017}). The \texttt{ARTEMIS} project has received funding from the European Research Council (ERC) under the European Union's Horizon 2020 research and innovation programme (grant agreement No 769130).  SEG and MEDR acknowledge support from {\it Agencia Nacional de Promoci\'on de la Investigaci\'on, el Desarrollo Tecnol\'ogico y la Innovaci\'on} (Agencia I+D+i, PICT-2021-GRF-TI-00290, Argentina). ASF acknowledges support from UKRI (ST/W006766/1).

\section*{Data availability}
Data for the \texttt{EAGLE} simulations used in this study (galaxy and halo catalogues) are publicly accessible via SQL queries from the following url: \url{http://icc.dur.ac.uk/Eagle/database.php}. Data from \texttt{ARTEMIS} simulations and codes developed for the analysis of this paper can be shared on reasonable request with the corresponding author.



\bibliographystyle{mnras}
\bibliography{refs} 




\appendix

\section{Models with varying SN and AGN feedback}
\label{sec:appendix}

\begin{figure*}
\includegraphics[width=1.44\columnwidth]{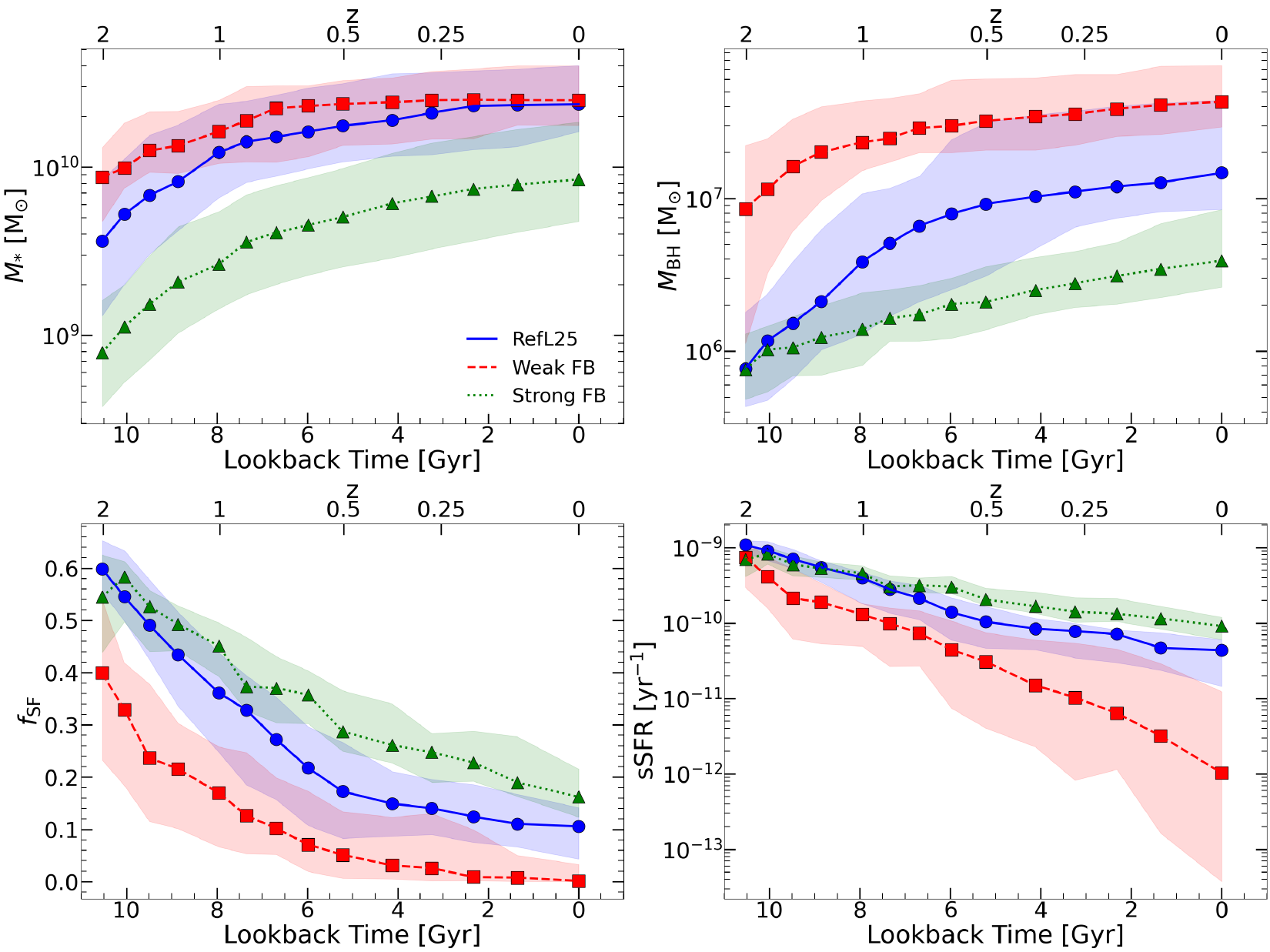}
\caption{Running medians for stellar and gaseous properties of MW-mass galaxies, $M_{\star}$, central $M_{\rm BH}$, $f_{\rm SF}$ and sSFR, in \texttt{EAGLE} runs with varying SN feedback. Blue lines represent the reference model, red lines the model with weak SN feedback and green lines the model with strong SN feedback. The shaded regions enclose the corresponding 25$^{\rm th}$ and 75$^{\rm th}$ percentiles.}
\label{fig:evo_properties_EAGLE_diff_FB}
\end{figure*}

\begin{figure*}
\includegraphics[width=1.44\columnwidth]{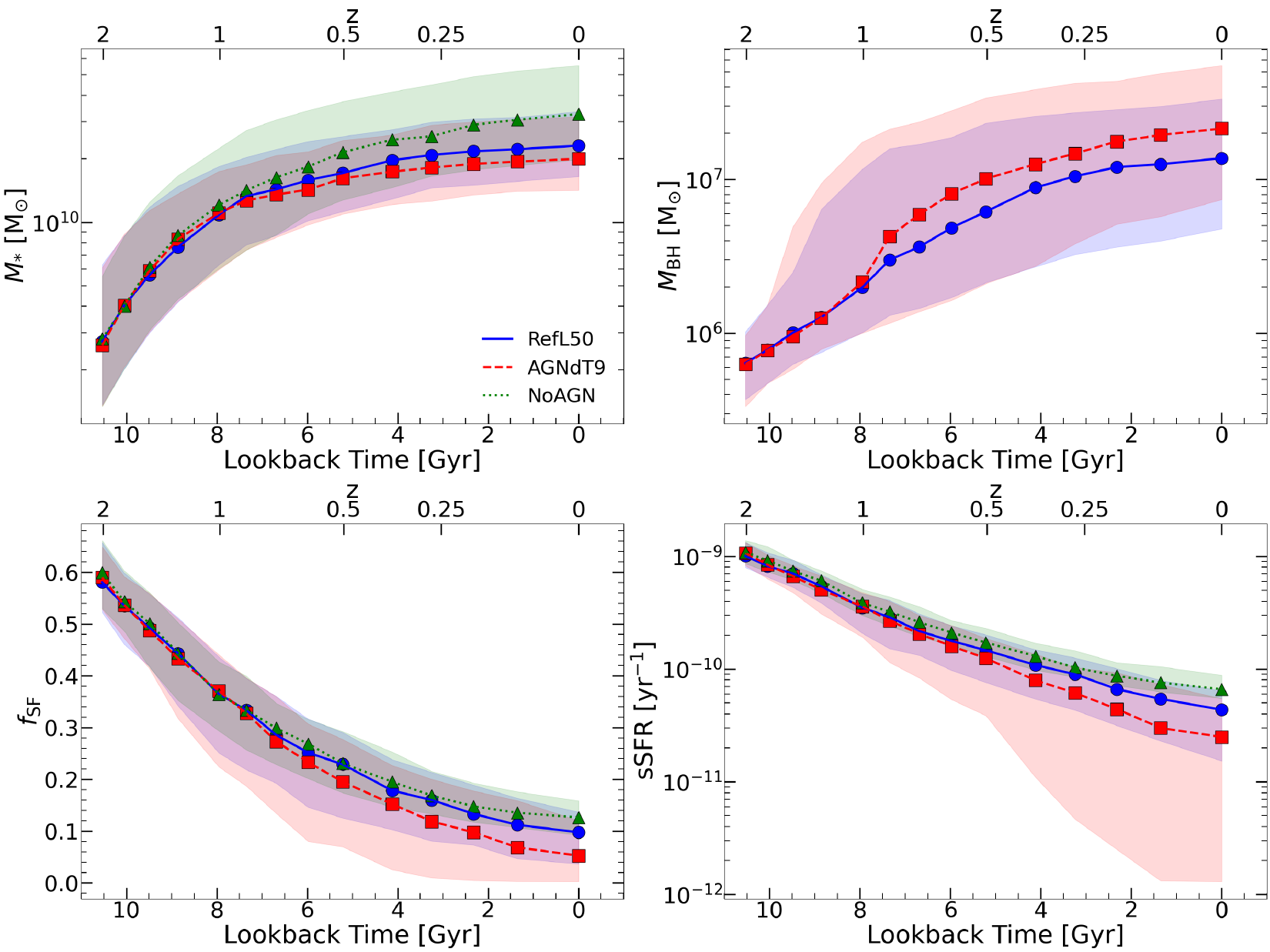}
\caption{Running medians as in Fig.~\ref{fig:evo_properties_EAGLE_diff_FB} but for the runs with different AGN feedback. Blue lines represent the reference model ${\Delta}T_{\rm AGN}=10^{8.5}~\mathrm{K}$, red lines the model with ${\Delta}T_{\rm AGN}=10^9~\mathrm{K}$ and green lines the model with no AGN.}
\label{fig:evo_properties_EAGLE_diff_AGN}
\end{figure*}

Here we compare the properties of $L^{\star}$ galaxies and their central BHs in several \texttt{EAGLE} runs with variations of SN and AGN feedback. As before, we select central galaxies of MW-mass, i.e., with present-day virial masses between $6.7\times10^{11}<M_{200,\mathrm{crit}}/\mathrm{M_{\odot}}<3.6\times10^{12}$. 

We use three \texttt{EAGLE} runs with varying SN feedback: i) a reference model with fiducial SN feedback; ii) a model with weak SN feedback, ``WeakFB''; and iii) one with strong feedback, ``StrongFB''. The last two models were obtained by scaling the subgrid parameter $f_{\mathrm{th}}$ by factors of $0.5$ and $2$, respectively. $f_{\mathrm{th}}$ represents the fraction of the energy budget in Type II SNe used for thermal heating (see \citealt{Crain2015}). All runs use the reference AGN feedback ($\Delta T_{\mathrm{AGN}}=10^{8.5}~\mathrm{K}$) and the same cosmological volume ($25$~cMpc on the side). Fig.~\ref{fig:evo_properties_EAGLE_diff_FB} shows the evolution of median stellar masses and gas fractions in MW-mass galaxies in these runs, as well of the masses of their central BHs.  Weaker SN feedback leads to earlier stellar mass assembly (by $z\simeq1$), producing a more rapid decline in both the star-forming gas fraction and the specific star formation rate (sSFR) over cosmic time, and consequently higher BH masses. This is expected, as weaker feedback allows more efficient gas cooling and faster gas consumption, resulting in earlier quenching and a larger quenched fraction by (z=0). In contrast, stronger SN feedback delays star formation, leading to lower stellar masses, higher sSFRs at late times, including at $z=0$, and lower BH masses.

Although neither of these scenarios is consistent with formation of MW-mass galaxies, the results highlight the sensitivity of galaxy and BH properties to this type of feedback. For example, changes in SN feedback efficiency by a factor of 2 lead to variations on an order of magnitude in $M_{\star}$ and $M_{\mathrm {BH}}$. Another interesting result is that the scatter in $M_{\mathrm {BH}}$ for the two non-fiducial SN models is very narrow, which is also inconsistent with the observed scatter in the $L^{\star}$ regime (see Fig.~\ref{fig:Mstar_MBH_relation}). 

We repeat the analysis for three \texttt{EAGLE} runs with varying AGN feedback: i) a model with the reference AGN feedback ($\Delta T_{\mathrm{AGN}}=10^{8.5}~\mathrm{K}$; 
also included in Figs.~\ref{fig:Mstar_MBH_relation} and \ref{fig:MBH_scaling_relations}); ii)  a model with increased AGN feedback, ``AGNdT9'' ($\Delta T_{\mathrm{AGN}}=10^{9}~\mathrm{K}$); and iii) one with no AGN feedback. These were run in a cosmological volume of $50$~cMpc and use the fiducial SN feedback (see \citealt{Schaye2015} for details). A higher $\Delta T_{\mathrm{AGN}}$ leads to more energetic feedback episodes, generally causing smaller radiative losses in the ISM, while a smaller $\Delta{T_{\rm AGN}}$ results in more intermittent AGN feedback.  Fig.~\ref{fig:evo_properties_EAGLE_diff_AGN} shows the co-evolution of galaxy stellar masses, gas fractions and sSFRs with BH masses for these runs. In this case, changes in AGN feedback result in modest variations in galaxy and BH properties, which suggests that AGN feedback is not the dominant process at this galaxy scale. However, its effect is more noticeable at later times $(z\lesssim1)$, when it facilitates the growth of BHs. Moreover, the model without an AGN does not produce any quenched galaxies (i.e., sSFR $<10^{-11}~{\rm yr}^{-1}$) and therefore cannot match the observations, particularly those of $L^{\star}$ galaxies with massive BHs. This implies that AGN feedback remains an essential subgrid physics ingredient for modeling galaxy evolution, even at the scale of MW-mass galaxies.


\bsp	
\label{lastpage}
\end{document}